\documentclass[fleqn,usenatbib]{mnras}

\usepackage{newtxtext,newtxmath}

\usepackage[T1]{fontenc}
\usepackage{ae,aecompl}
\renewcommand{\arraystretch}{1.7}

\usepackage{graphicx}	
\usepackage{amsmath}	
\usepackage{amssymb}	
\usepackage[rightcaption]{sidecap}
\usepackage{caption}
\usepackage{float}
\usepackage{gensymb}
\usepackage{subfig}

\title[Radiative components for QPOs]{Identifying the radiative components responsible for Quasi-Periodic Oscillations of black hole systems}

\author[Garg et al.]{
Akash Garg,$^{1}$\thanks{E-mail: akashgarg\_16@yahoo.co.in}
Ranjeev Misra,$^{2}$
Somasri Sen$^{1}$
\\
$^{1}$Department of Physics, Jamia Millia Islamia, Jamia Nagar, New Delhi-110025, India\\
$^{2}$Inter-University Centre for Astronomy and Astrophysics, Ganeshkhind, Pune-411007, India\\
}

\date{Accepted XXX. Received YYY; in original form ZZZ}

\pubyear{2019}

\begin{document}
\label{firstpage}
\pagerange{\pageref{firstpage}--\pageref{lastpage}}
\maketitle

\begin{abstract}
While the dynamical origin of the variability observed in Galactic Black hole systems, such as quasi-periodic oscillations (QPOs), are still a matter of debate, insight into the radiative components responsible for such behaviour can be obtained by studying their energy dependent temporal behaviour. In particular, one needs to ascertain which variations of the parameters of the best fit time-averaged spectral components reproduce the observed energy dependent fractional r.m.s and time-lags. However, to obtain meaningful interpretation, the standard spectral component parameters have to be recast to physically relevant ones. Then, the energy dependent temporal variations that their fluctuations will cause, needs to be predicted and compared with observations. In this work, we describe a generic method to do this and apply the technique to the $\sim$ 3-4 Hz QPOs observed in the black hole system GRS 1915+105 as observed by {\it AstroSat} where the time-averaged spectra can be represented by emission from a truncated disk and a hot thermal Comptonizing coronae in the inner regions. We find that the QPOs and their harmonic can be explained in terms of
correlated local accretion rate variations in the disk, the truncated disk radius, the optical depth and the heating rate of the coronae with time-delays between them. We highlight the potential of such techniques to unravel the radiative process responsible for variability using high quality spectral and temporal data from  {\it AstroSat} and {\it NICER}.
\end{abstract}




\section{INTRODUCTION}

The Black hole X-ray binaries (BHXB) have drawn considerable interest owing to their large X-ray luminosity and variability on time scales of years to milliseconds. These systems are modelled as a Black hole accreting matter from a companion star as it evolves and fills its Roche lobe or directly from the companion's wind. The accreting matter forms a disk around the Black hole which radiates  X-rays by extraction of energy from the gravitational potential. These sources provide a window to explore general relativity in the strong regime and to put constraints on the properties of black holes. It has been long believed that a proper interpretation of the rapid variability observed in this sources will be one
of the primary methods for providing such information.\\

Rapid variability is quantified using the Power Density Spectrum (PDS), which is the amplitude squared of fourier transform of light curves generated to look for characteristic frequencies. The PDS of many BHXB have various features from broad structures called noise to narrow peaks
known as Quasi periodic oscillations(QPOs)(e.g. \citet{Motta S.E.(2016)}). Various detections of QPOs in BHXB have suggested a classification on the basis of frequency range- Low frequency QPOs(LFQPOs) in the 0.1-30 Hz and high frequency QPOs(HFQPOs) in 40-450 Hz(\citet{Remillard & McClintock(2006)}). LFQPOs  have been categorized further into Types A, B and C on the basis of intrinsic properties such as  centroid frequency, width, energy dependence of variability etc. They are detected in different spectral states of a source, such as low-hard, hard-intermediate, or soft-intermediate states which are typically identified using  hardness-intensity diagram \citep{Wijnands et al.(1999),Homan et al.(2001),Remillard et al.(2002)}. After analyzing multiple observations for the detection of HFQPOs \citet{Belloni et al.(2012)} found them in sources like XTE J1550-564 and GRO J1655-40.\\

Though there have been large number of observations of QPOs, their origin is not properly understood. There exist different models that explore
the diverse physical scenario that may explain the phenomenon in BHXB. Some of these models use the Lense-Thirring precession where a misaligned accretion disk causes differential precession of particle orbits \citep{Fragile et al.(2007),Ingram et al.(2009)}. Besides this, \citet{Tagger et al.(1999)} and \citet{Rodriguez et al.(2002a)} provided another explanation in the form of accretion-ejection instability which involves transportation of energy and momentum from inner regions of disk towards co-rotation radius and then emitted towards hotter parts of
the disk, resulting in standing waves giving rise to QPOs. \citet{Titarchuk et al.(2000)} discussed the possibility of disk oscillations due to gravitational interaction between the compact object and the disk. Here, the gravity acts as a restoring force for displacements in the disk leading to modes of frequencies corresponding to the QPOs in BHXB. An alternative interpretation is given in the shock oscillation model(\citet{Chakrabarti & Manickam(2000)}) where shocks located at the edge of Keplerian disk can cause oscillations in the accretion flows producing the characteristic frequencies related to LFQPOs. In truncated disk models, the inner disk is hidden/absent due to the presence of hot inner flow which tends to undergo vertical precession, resulting in QPOs (\citet{Axelsson et al.(2014)}). Such theoretically distinct models need to be validated or ruled out by observations. And at present there is still a lack of consensus on the most favourable one.\\

Typically, the approach has been to identify the QPO frequency with that corresponding to a theoretical prediction. An alternate approach is to identify the radiative process that gives rise to the QPO and thereby getting insight into the possible underlying dynamic process. Indeed, correlations between the QPO frequency and time averaged spectral parameters indicate a strong connection between the QPO phenomenon and the geometry and radiative state of the inner regions. For instance, \citet{Belloni et al.(2000)} showed that the characteristics of the QPOs depend on the spectral states for the black hole system GRS 1915+105. For one of the soft luminous state (State A) the PDS shows weak and broad peaks around 6-8 Hz whereas for another one (state B), relatively strong and narrow peaks at 1-3 Hz are observed. On the other hand for the hard state at low luminosity, where the spectrum was dominated by a hard power-law component, the system exhibited strong 0.5-10 Hz QPOs. More importantly, it has been shown that the 1.7-3.0 Hz QPO frequency correlates tightly with spectral component parameters(\citet{Bhargava et al.(2019)}). \citet{Misra et al.(2013)} explained the alternating lags observed for 67 mHz QPO in GRS 1915+105 and its harmonics by considering variations on spectral parameters with time delays. \citet{Mir et al.{2016}} worked on the same lines to model the energy dependent behaviour for the heartbeat oscillations of frequency 20.0-80.0 mHz in GRS 1915+105. Their model named "DROID" assumed that there is a delayed response of the inner disk radius to variations in the accretion  rate. In a different approach, \citet{Uttley et al.(2011)} attributed the soft lags in GX 339-4 to X-ray heating in which variations in powerlaw and disk have a positive or negative lag depending on the time scales of variability.\\

These results were primarily based on data from the {\it Rossi X-ray timing Experiment} (RXTE). Indeed, RXTE data has provided detailed energy dependent information for QPOs \citep{Reig et al.(2000),Pahari et al.(2013)}. The Large Area X-ray Proportional counter (LAXPC) on board the observatory {\it AstroSat} \citep{Yadav et al.(2016a),Agrawal et al.(2017)} extends the high energy range of RXTE of 3-30 keV to about 3-80 keV. LAXPC data has revealed the complex nature of the energy dependent properties of the 0.1-10 Hz QPO for GRS 1915+105 \citep{Yadav et al.(2016b),Divya et al.(2019)}. A unique advantage of {\it AstroSat} is that the Soft X-ray Telescope (SXT) on board provides
simultaneous spectral coverage in the soft energy band, thereby allowing for broad band spectral fitting in the 0.3-80 keV range(\citet{Singh et al.(2016)}). The advantage of
such broad band spectral coverage and timing information was demonstrated by \citet{Bari et al.(2019)} where they invoked a one zone stochastic propagation model to explain the energy dependent fractional r.m.s. 

Thus, it is important to develop techniques that can test model's prediction of timing properties based on broad band spectral information. In the works mentioned above, specific techniques were developed to address the timing information from  the specific spectral model used to fit the data.
It is desirable to have a generic technique where the timing properties related to a spectral model used can be obtained. This will allow for testing which spectral model is favoured by the timing observations.\\

In this work, we outline such a technique and as an example test it on a QPO observed by  AstroSat/LAXPC data for the black hole system GRS 1915+105. To the linear order the temporal properties can be obtained by numerically differentiating model spectral components and using them to estimate variations of spectral parameters that can explain the timing results. However, a critical aspect is that typically the parameters of a spectral component are often chosen to be such that they are non-degenerate and are not necessarily physical ones. Thus, for temporal analysis, the spectral model needs to be transformed such that it is represented by physical quantities, whose variations can then be used to predict timing properties. In this work, we show how this maybe done for a simple but important case of a spectrum described by a multi-colored truncated disk and a thermal plasma which Comptonizes photons from the disk.\\

In the next section, we describe the technique while in Section~\ref{sec:Obser} we apply it to a QPO observed by AstroSat/LAXPC for GRS 1915+105.
In the last section we discuss the results.

\section{MODELLING THE ENERGY DEPENDENT TEMPORAL PROPERTIES}
\label{sec:numer}

The time averaged spectra of Black Hole binaries are often dominated by two spectral components, one arising from a multi-coloured disc and the other a thermal Comptonized spectrum from a hot corona. Typically the spectra are analyzed with the software XSPEC using models defined within. For example, the disc emission is often represented by the XSPEC function "diskbb", while the thermal Comptonization is represented by "nthcomp". The best fit parameters for these models are reported with errors. Energy dependent time-lag and r.m.s can be used to determine which of these model parameters are responsible for the source variability.\\

The parameters of model functions used to fit spectra, are usually chosen such that during fitting, the parameters are independent and hence reliable best fit parameters with errors can be estimated. Thus, the chosen parameters may not have a one to one correspondence with physically relevant ones. For a standard multi-coloured accretion disk, the relevant physical parameters that may vary during an observation are the accretion rate and the inner disk radius under the assumption that the inclination angle and the color factor do not vary or vary insignificantly. The XSPEC model function "diskbb" has two parameters: the normalization which is directly related to the inner radius of the disk and the inner disk temperature $kT_{in}$ which is related to the accretion rate and the inner disk radius. Thus, variations in the normalization correspond to inner radii changes, while $kT_{in}$ variations (with normalization constant) would correspond to accretion rate changes. Thus, in this simple case, it is rather straightforward to relate the parameter variations of the XSPEC model to physically relevant ones.\\

The situation is less straightforward for more complex models like thermal Comptonization. The XSPEC model "nthcomp" has four parameters. The inner disk temperature $kT_{in}$ is used to characterize the input seed photon and is the same as that of the model "diskbb". The other parameters are the temperature of the Comptonizing medium $kT_e$ and the asymptotic power-law index $\Gamma$. Moreover, there is a normalization parameter which is the value of the photon spectrum at 1 KeV, since internally the model code re-normalizes the spectrum such that its value at 1 keV is unity. The asymptotic power-law index $\Gamma$ is a function of the temperature $kT_e$ and optical depth $\tau$\citep{Zdziarski et al.(1996),Zycki et al.(1999),Wilkins et al.(2015)},
\begin{equation}
  \Gamma = [9/4+(3m_ec^2)/(kT_e((\tau+3/2)^2-9/4))]^{1/2}-1/2
  \label{eq:gam}
\end{equation}
\noindent
Internally the model code uses $\tau$ for spectral computation after obtaining its value using Equation \ref{eq:gam}. These parameters have been chosen to facilitate spectral fitting, and their correspondence to physically relevant parameters are not straight forward. The physical quantities or parameters which may vary are the normalization $N_{dbb}$ (related to the inner radius) and inner disk temperature $kT_{in}$ of the disk emission, the fraction $f$ of the disk photons which enter the Comptonizing region, the optical depth $\tau$ and the heating rate of the corona $\dot H$. It is the heating rate of the corona, together with other parameters which determines its temperature $kT_e$. Specifically, 
\begin{equation}
    \dot{H} = \int
    E(F_{c}(E,kT_e,\tau)-F_{inp}(E)) dE
    \label{eq:5}
\end{equation}
\noindent
where $F_c$ is the photon flux from the corona, $F_{inp} (E) = f*F_{d}$ is the seed photon flux and $F_d (E,kT_{in})$ is the disk flux.\\ 

\begin{figure*}

\centering
\includegraphics[width=0.5\textwidth,height=5cm]{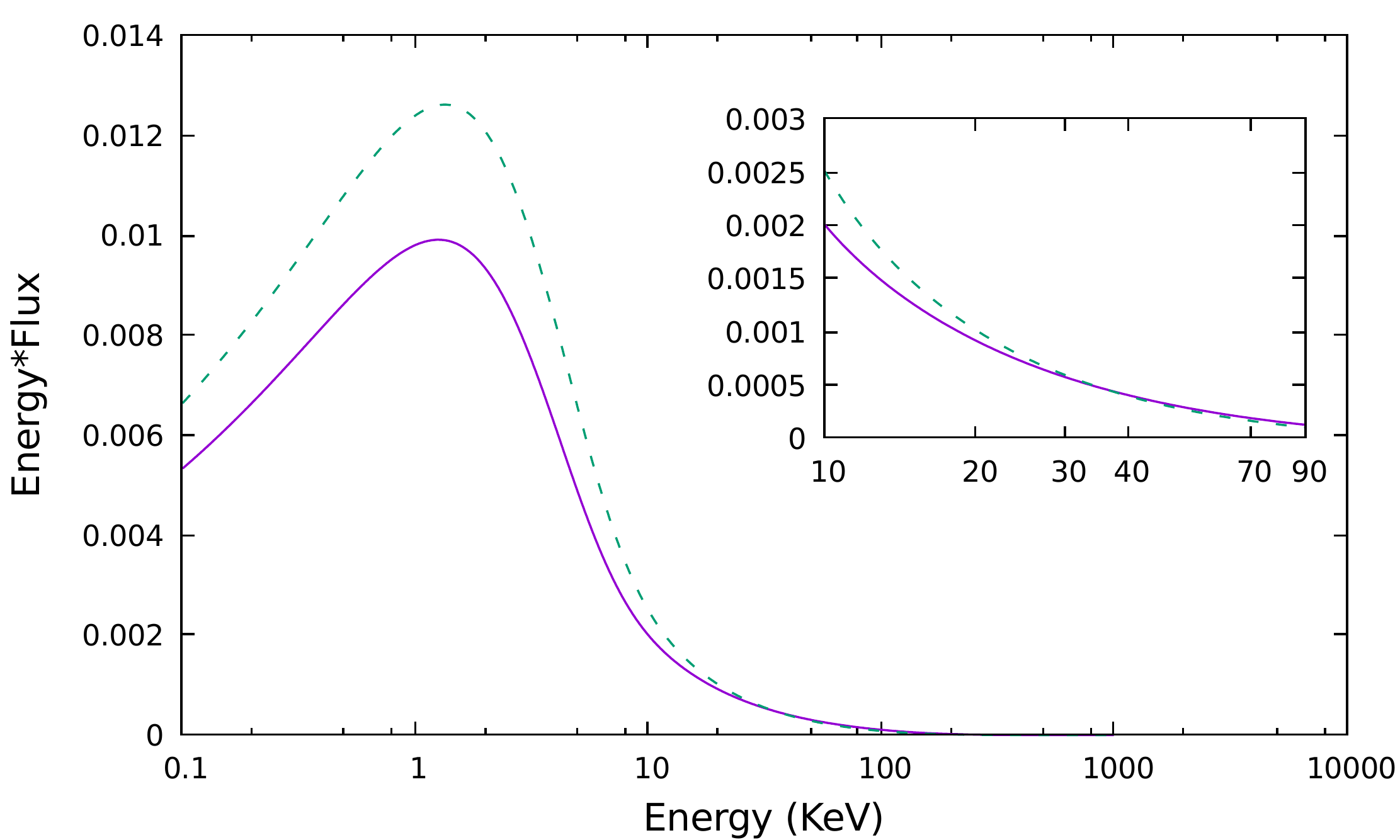}\hfill
\includegraphics[width=0.5\textwidth,height=5cm]{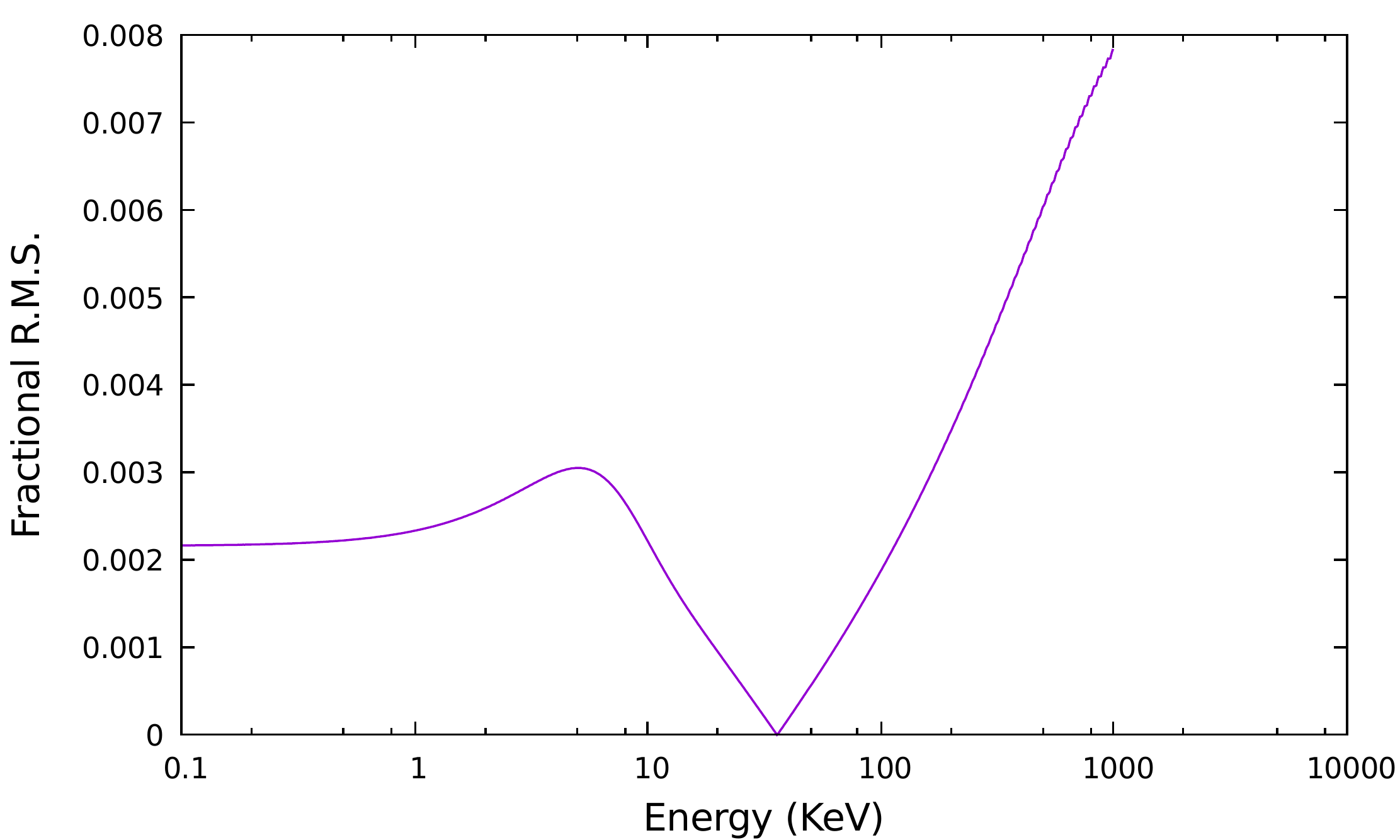}

\caption{Left panel represents the spectrum for two different seed photon temperature with pivoting happening at energy ($\sim$35 KeV) which is also reflected in the adjacent panel where fractional rms spectra shows non-monotonic behaviour around the pivot.}
\label{fig:model1}

\end{figure*}

\begin{figure*}

\centering
\includegraphics[width=0.5\textwidth,height=5cm]{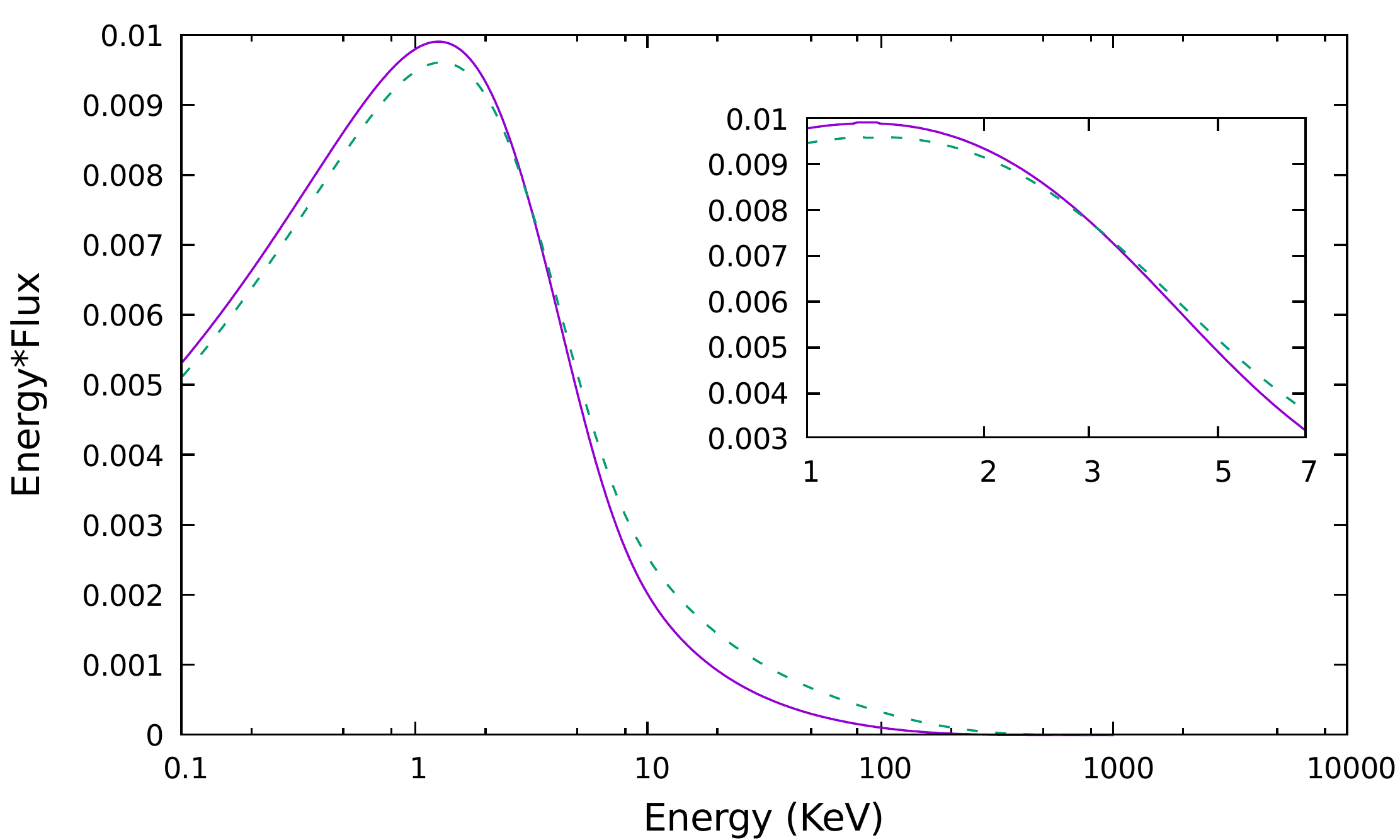}\hfill
\includegraphics[width=0.5\textwidth,height=5cm]{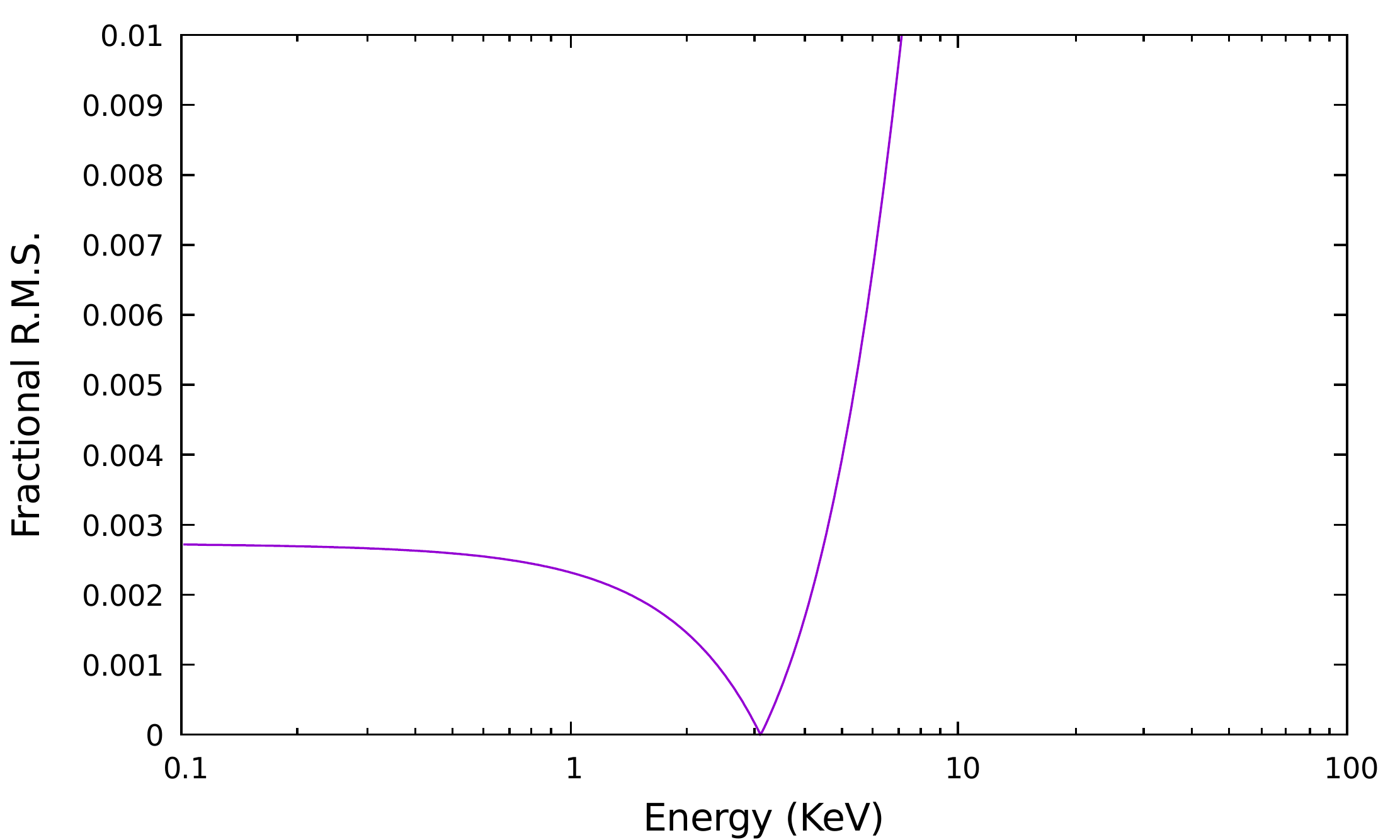}

\caption{Left panel represents the spectrum for two different heating rate. The spectra pivots at energy ($\sim $ 3.0 KeV), observed as well in the adjacent panel showing fractional rms spectra.}
\label{fig:model2}

\end{figure*}

\begin{figure*}

\centering
\includegraphics[width=0.5\textwidth,height=5cm]{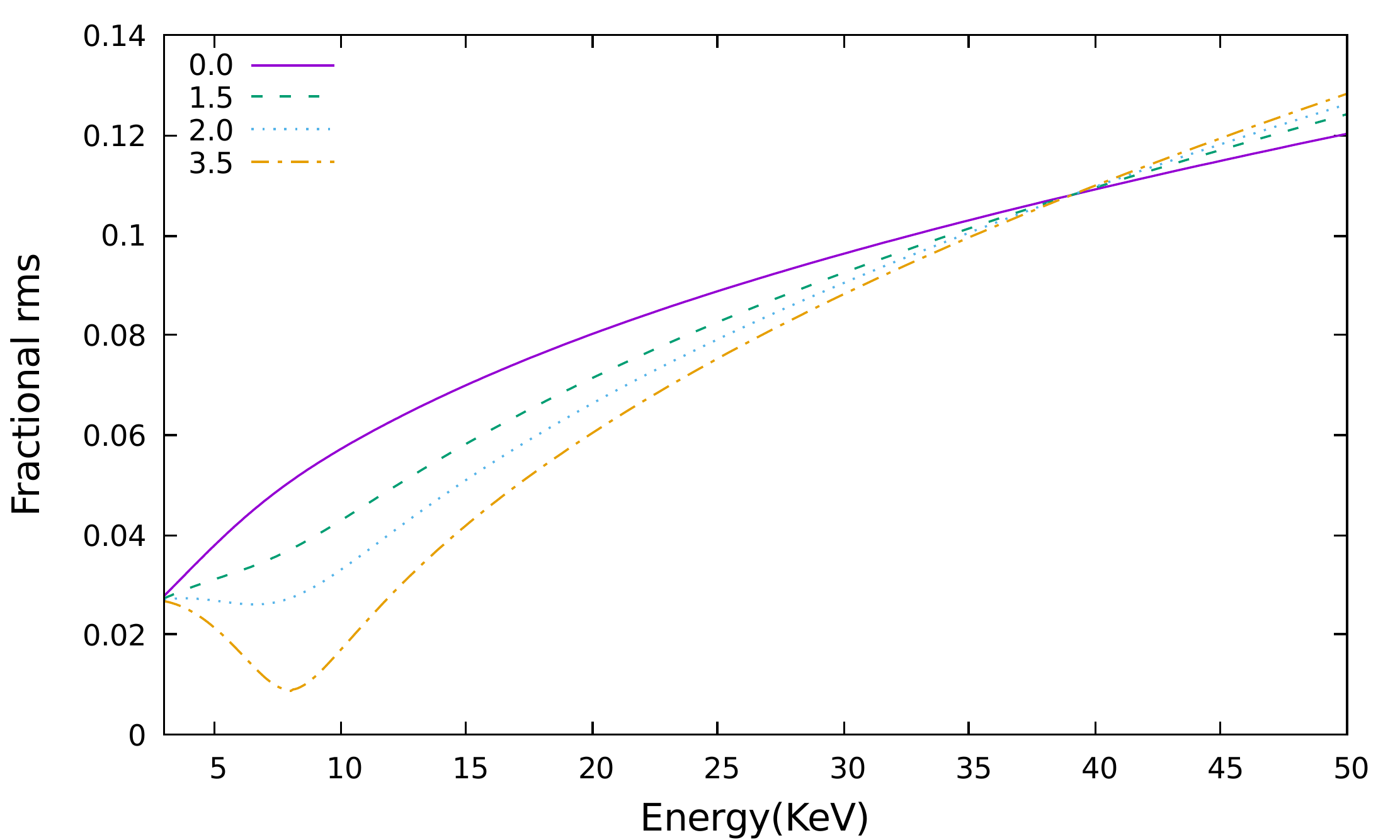}\hfill
\includegraphics[width=0.5\textwidth,height=5cm]{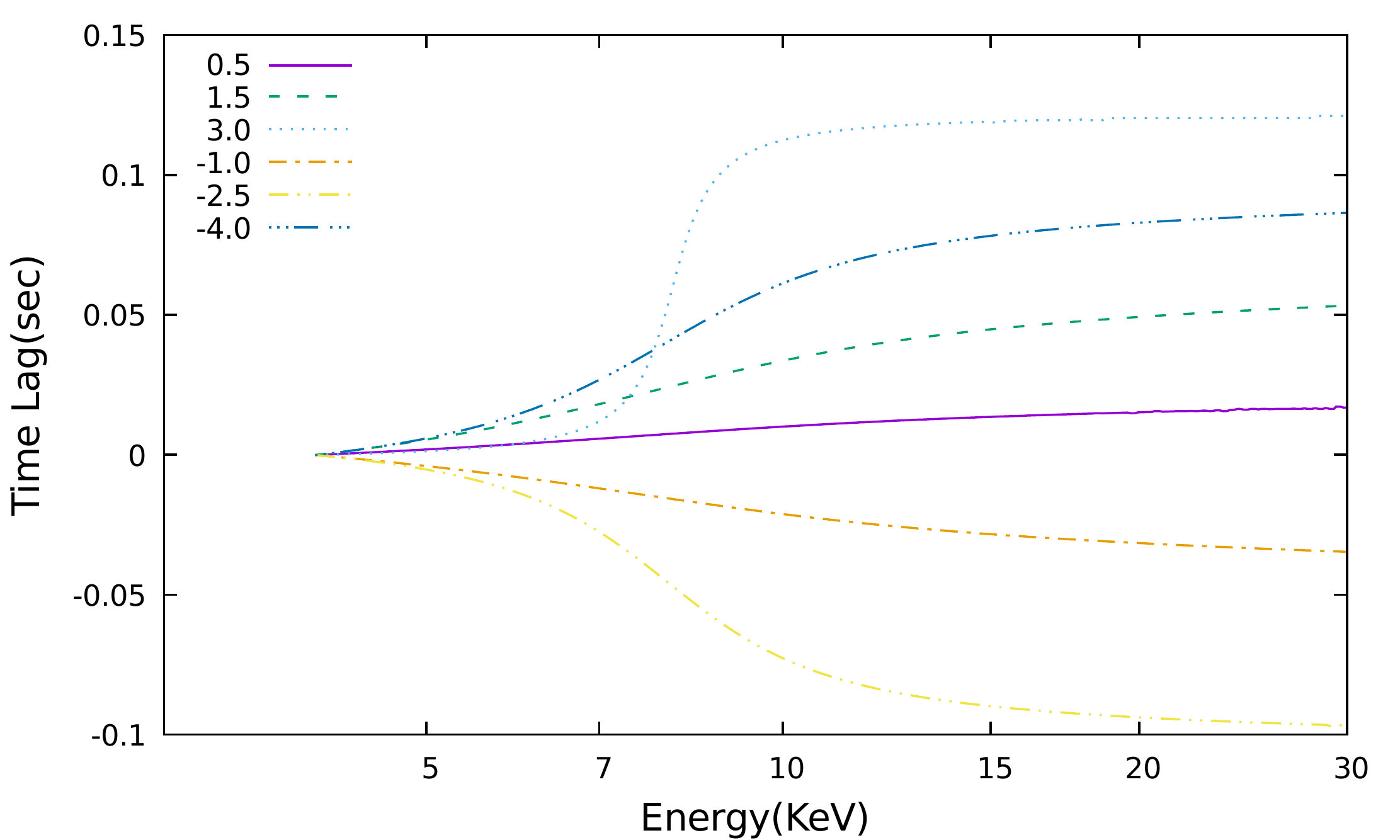}

\caption{Left and right panels present the fractional rms and time lag versus energy for different values of phase difference between $ \Delta kT_{in}$ and $ \Delta \dot{H}$, keeping all other parameters constant at some fiduciary values.}
\label{fig:model3}

\end{figure*}

In order to quantify the physical parameters responsible for the variability, one needs to recast the spectral model "nthcomp" as a function of those parameters. Hence, we modified the model code such that the resultant spectrum $F_c^\prime$ is a function of $kT_{in}$, $f*N_{dbb}$, $\dot{H}$ and $\tau$ instead of $kT_{in}$, $N_c$, $kT_{e}$ and $\Gamma$. To do so we first remove the re-normalization of the spectrum to be unity at 1 keV and instead make the normalization of the seed photon spectrum to be $f*N_{dbb}$. In practise, we keep the normalization of the input seed photon to be unity and multiply the computed spectrum by $f*N_{dbb}$, which is equivalent to having the input spectrum to have that normalization. Next, instead of $\Gamma$, the optical depth $\tau$ is directly used as an input parameter. Finally the input parameter $\dot{H}$, is used to compute the electron temperature $kT_e$ from Equation \ref{eq:5}. This is done iteratively i.e. by refining guess values of $kT_e$ till the integration is equal to the input value of $\dot H$.\\

To reiterate, by this technique one can recast the disk and Comptonization spectral components having parameters $N_{dbb}, kT_{in}, N_c, kT_e$ and $\Gamma$ to those having parameters $N_{dbb}, kT_{in}, fN_{dbb}, \dot H$ and $\tau$. We note that the spectral fitting should be done using the earlier set of parameters and hence one would obtain the time averaged spectral parameters $N_{dbb}^o, kT_{in}^o, N_c^o, kT_e^o$ and $\Gamma^o$, which are then converted to $N_{dbb}^o, kT_{in}^o, fN_{dbb}^o, \dot H^o$ and $\tau^o$ for computing the energy dependent variability as described below.\\

To the first order the variation of the spectrum to changes in spectral parameters can be estimated by
\begin{equation}
  \Delta F (E) = \sum_{j=1}^M \frac{\partial F (E)}{\partial \alpha_j} \Delta \alpha_j
  \label{DeltaSE}
\end{equation}
\noindent
where $F(E)$ is the steady state spectrum and $\alpha_j$ are the spectral parameters while $\Delta \alpha_j$ are the small variations of the parameter values, which in general may be complex numbers and $M$ is the number of parameters. The fractional r.m.s. as a function of energy
will then be given by $|\Delta F(E)|/F(E)$ and the phase-lag between photons of energy $E$ and a reference energy $E_{ref}$, would be the argument of ${\Delta F(E_{ref})}^*\Delta F(E)$. The partial derivatives $ \frac{\partial F (E)}{\partial \alpha_j} $ can be computed numerically by
\begin{equation}
  \frac{\partial F (E)}{\partial \alpha_j} \sim \frac{ F(E,\alpha_j^o+\delta \alpha_j)-F(E,\alpha_j^o)}{\delta \alpha_j}
\end{equation}
\noindent
where $\delta \alpha_j << \alpha_j$ is a small variation and $F(E,\alpha_j^o+\delta \alpha_j)$ is computed with all the rest of the
parameters being held constant to their steady state values.\\

One of the variations $\Delta \alpha_j$ can be taken as reference (and hence limited to be positive real), the other in general can have a phase. Hence the number of parameters that are required to describe the r.m.s and phase lag as a function of energy are $2M-1$, where $M$ is the number of independent physical parameters used in the time-averaged spectral fitting. Since often $2M-1$ maybe a large number of parameters, one needs to make a choice of which physical parameter to vary and to introduce a phase difference only between some of them.\\

To illustrate the effect of parameter variation on the spectrum, we show in Figure~ \ref{fig:model1} and \ref{fig:model2}, the spectral shape change and the corresponding r.m.s versus energy for variation in the input photon temperature, $kT_{in}$ and the coronal heating rate $\dot H$. The spectral parameters used for this illustration are $kT_{in}=1.2 KeV, kT_e=100 KeV$ and $\Gamma= 1.9$ which correspond to the physical parameters, $kT_{in}=1.2 KeV, \dot{H} = 5.0 KeV/cm^3/sec$ and $\tau = 1.0 $. Variability in the inner disk temperature which corresponds to change in the disk accretion rate makes the spectrum pivot around 35 keV, while coronal heating variation leads to a pivot around 3 keV. The exact values of these pivot points depend on the spectral parameters assumed, but qualitatively the pivot point is at high energies when the inner disk temperature is varied while it is at relatively low energies when the heating rate is varied. When a single spectral parameter is varied, there is no energy dependent time-lag between the photons, except for a trivial 180$^\degree$ abrupt phase shift across the pivot point. The combination of these variations can lead to different types of energy dependent r.m.s and time-lag behaviour as illustrated in Figure~ \ref{fig:model3}. Here both $kT_{in}$ and $\dot{H}$ have been varied with the ratio $\delta kT_{in}/\delta \dot{H} = 0.10 cm^3 sec$. The energy dependence of the r.m.s and time-lag are plotted for a set of phase difference between the two variables. Note that in this case there is no pivot point i.e. the fractional r.m.s does not go to zero for any energy. It is also important to note that the r.m.s versus energy depends significantly on the phase difference. The scenario becomes more complex upon introduction of the possibility of a third spectral parameter varying with the functional form of the energy dependent photon r.m.s and time-lag depending on the amplitudes of the variations and the time-lag between them. Thus, such combination of variations of physical parameters can be tested by comparison with observations, like in the example given in the next section.

\section{COMPARISON WITH OBSERVATIONS OF GRS 1915+105}
\label{sec:Obser}

\citet{Divya et al.(2019)} have reported the timing analysis of AstroSat/LAXPC observation of the black hole X-ray binary GRS 1915+105, on 28th March 2017. They reported a strong QPO whose frequency varied over time as the source made transition between $\chi$ and $\rho$ class and presented the energy dependent r.m.s and time lags. We have chosen this observation to illustrate how one can attempt to identify the radiative component that is responsible for the variation. As reported by \citet{Divya et al.(2019)}(see also \citet{Yadav et al.(2016b)}) prominent broad peaked features can be seen corresponding to the frequencies $\sim$3.5 Hz, $\sim$3.78 Hz and $\sim$3.91 Hz in the different sections of the data. For completeness, we re-analyzed the $\sim$ 13 Ksec of the whole data to obtain the energy dependent QPO properties. We divided the data into three sections and performed spectral and temporal modelling separately for each of them and refer to them as section I, II and III.\\

We calculated the fractional rms and time lags in different energy bands at these distinct QPO frequencies. We first generated the PSD for a particular energy band and fitted using multiple lorentzian components. The normalization of the fundamental or harmonic component is the integration of the PSD over the frequency range and therefore its square root yielded the normalized root mean square variability. The LAXPC subroutine, {\t laxpc\_find\_freqlag}  calculates the phase of the average cross power spectrum of two uniformly sampled light curves and then dividing it by $2\pi f$ to fourier time lags(\citet{Nowak et al.(1999)}). We used it to generate time lags in different energy bands at the QPO frequencies and their harmonics with respect to a reference energy band of  3.0-5.0 KeV.\\

The time averaged spectra in the energy range 3.0-30.0 KeV requires a combination of XSPEC components "diskbb", "nthcomp" and "gaussian" to model the disk blackbody, thermal comptonized component and iron line emission respectively as shown by \citet{Yadav et al.(2016b)}. However, for simplicity, we choose to fit the spectra using only the disk and thermal comptonized components.  A systematic error of 3$\%$ was added to account for uncertainties in the response calibration. The absorption column density $N_H$ was fixed at $\sim$ $4\times10^{22}$ $cm^{-2}$(\citet{Blum et al.(2009)}) and represented by the XSPEC component $\textit{Tbabs}$. The temperature of the thermal plasma could not be constrained and was fixed at $100$ keV.\\

\begin{figure}
\includegraphics[width=0.5\textwidth,height=6cm]{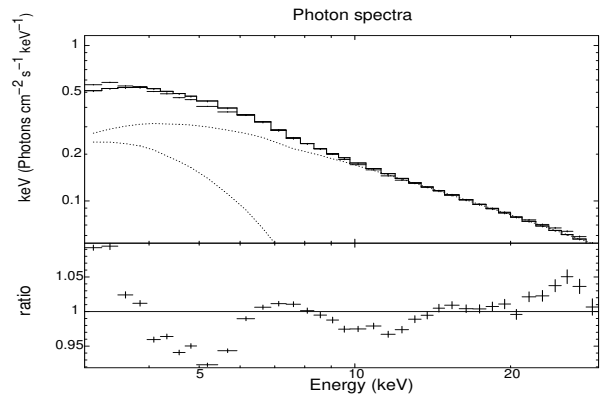}
\caption{Shows Photon spectra of GRS 1915+105 in 3.0-30.0 KeV range for section I of the data as observed by LAXPC10(Observation Id. 9000001116). The top panel shows data fitted with XSPEC model: Tbabs x (diskbb+nthcomp) with solid thick line representing the resultant model. The bottom panel gives details of the ratio of data and model. }
\label{fig:spectra}
\end{figure}

As an example Fig.~\ref{fig:spectra} shows the source's unfolded spectrum for section I. The bottom panel showing the ratio of the data to the model, which clearly shows the omission of the Iron line at $\sim$ 6.4 keV and that this will give rise to about $\sim$5\% error in the spectral fitting. Table~\ref{tab:spectra} lists the details of fitted spectral parameters.\\

\begin{table*}
    \centering
	\caption{Best fit spectral parameters for three different sections of data.}
	\label{tab:spectra}
	\begin{tabular}{c @{\hskip 1cm}c c@{\hskip 1cm} c @{\hskip 1cm}c} 
		\hline \hline
		 & & & Sections &\\
		\hline    
		Parameters & Description/Unit & I & II & III\\
		\hline
		$T_{in}$ & Inner disk temperature(KeV) & $1.188^{+0.005}_{-0.005}$ & $1.219^{+0.006}_{-0.006}$ & $1.231^{+0.006}_{-0.006}$\\
		$\Gamma$ & Photon Index                & $2.085^{+0.006}_{-0.006}$ & $2.106^{+0.008}_{-0.008}$ & $2.126^{+0.008}_{-0.008}$\\
		$\tau$ & Optical depth                 & $0.889^{+0.004}_{-0.004}$ & $0.871^{+0.006}_{-0.006}$ & $0.855^{+0.006}_{-0.006}$\\
		$\dot H$ & Heating rate($ keV/cm^3$/s) & $5.249^{+0.023}_{-0.080}$ & $4.924^{+0.058}_{-0.094}$ & $4.807^{+0.045}_{-0.060}$\\		
		$N_{dbb}$ & Normalization (diskbb)     & $94.056^{+5.507}_{-5.321}$ & $101.711^{+3.478}_{-3.383}$ & $111.822^{+3.603}_{-3.508}$\\
		$R_{in}$ & Apparent inner disk radius(kms) &  $11.795^{+0.340}_{-0.338}$ & $12.265^{+0.207}_{-0.202}$ & $12.861^{+0.205}_{-0.200}$\\
		$Flux_{total}$ & Unabsorbed total flux(x $10^{-9}$ ergs/s/$cm^{2}$) & $7.695^{+0.066}_{-0.066}$ & $7.765^{+0.064}_{-0.063}$ &  $8.056^{+0.073}_{-0.072}$\\
		$Flux_{disk}$ & Disk flux(x $10^{-9}$ ergs/s/$cm^{2}$) & $1.443^{+0.063}_{-0.062}$ & $1.656^{+0.066}_{-0.064}$ &  $1.927^{+0.072}_{-0.071}$\\
		$Flux_{disk}/Flux_{total}$ & Ratio of Disk flux and total flux & $0.187^{+0.006}_{-0.006}$ & $0.213^{+0.006}_{-0.006}$ & $0.239^{+0.006}_{-0.006}$\\
		$\chi^2$/dof & Reduced chi-square & 28.95/35 & 32.47/35 & 32.44/35\\
		\hline
	\end{tabular}
	
	Note: \textbf{XSPEC} model: Tbabs x (diskbb+nthcomp). $KT_e$, electron temperature was fixed at 100 KeV. $Flux_{total}$ and $F_{disk}$ were estimated in the energy range 3.0-30.0 KeV using cflux in \textbf{XSPEC}. The normalization $N_{dbb}$ of \textit{diskbb} is related to apparent inner disk radius $R_{in}$ by $N_{dbb} = (\frac{R_{in}}{D_{10}})^2cos\theta $ where $R_{in}$ is in kms and $D_{10}$ is distance to the source taken in units of 10 Kpc and $\theta$ is the inclination angle of the disk. All errors are at the 2$\sigma$ level.
\end{table*}

Having fitted the spectra using the XSPEC routines "diskbb" and "nthcomp" with the original parameters, we transform the parameters to physically relevant ones as discussed in the previous section and refit the spectra. Since the electron temperature $kT_e$ was fixed in the initial fitting, we choose the optical depth $\tau$ to be fixed at $0.889$ as determined from the photon index, $\Gamma$ and $kT_e$. Thus we obtained a value of the heating rate $\dot H = 5.249^{+0.023}_{-0.080} $ $keV/cm^3$/s.\\

\begin{figure*}

\centering
\includegraphics[width=0.33\textwidth,height=4.8cm]{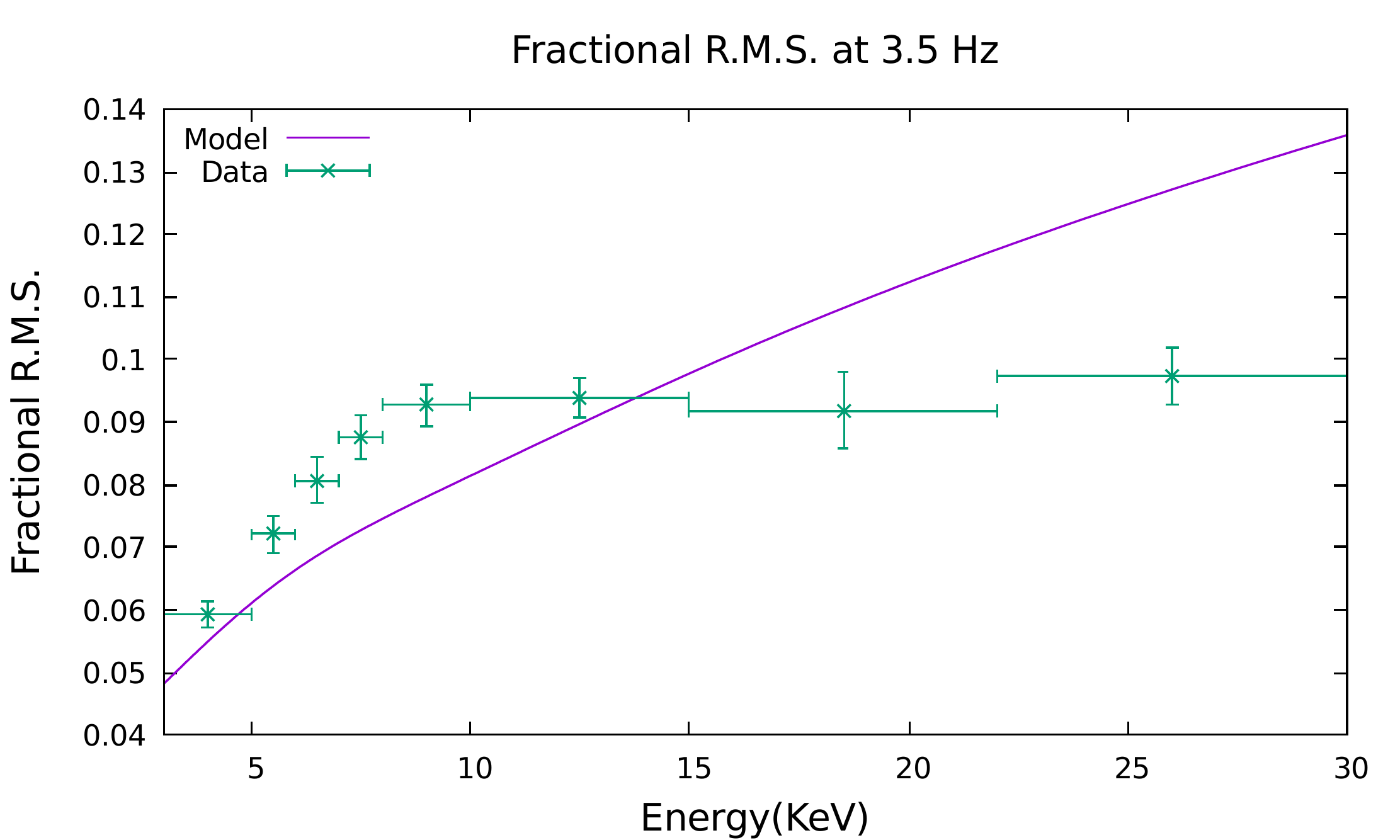}\hfill
\includegraphics[width=0.33\textwidth,height=4.8cm]{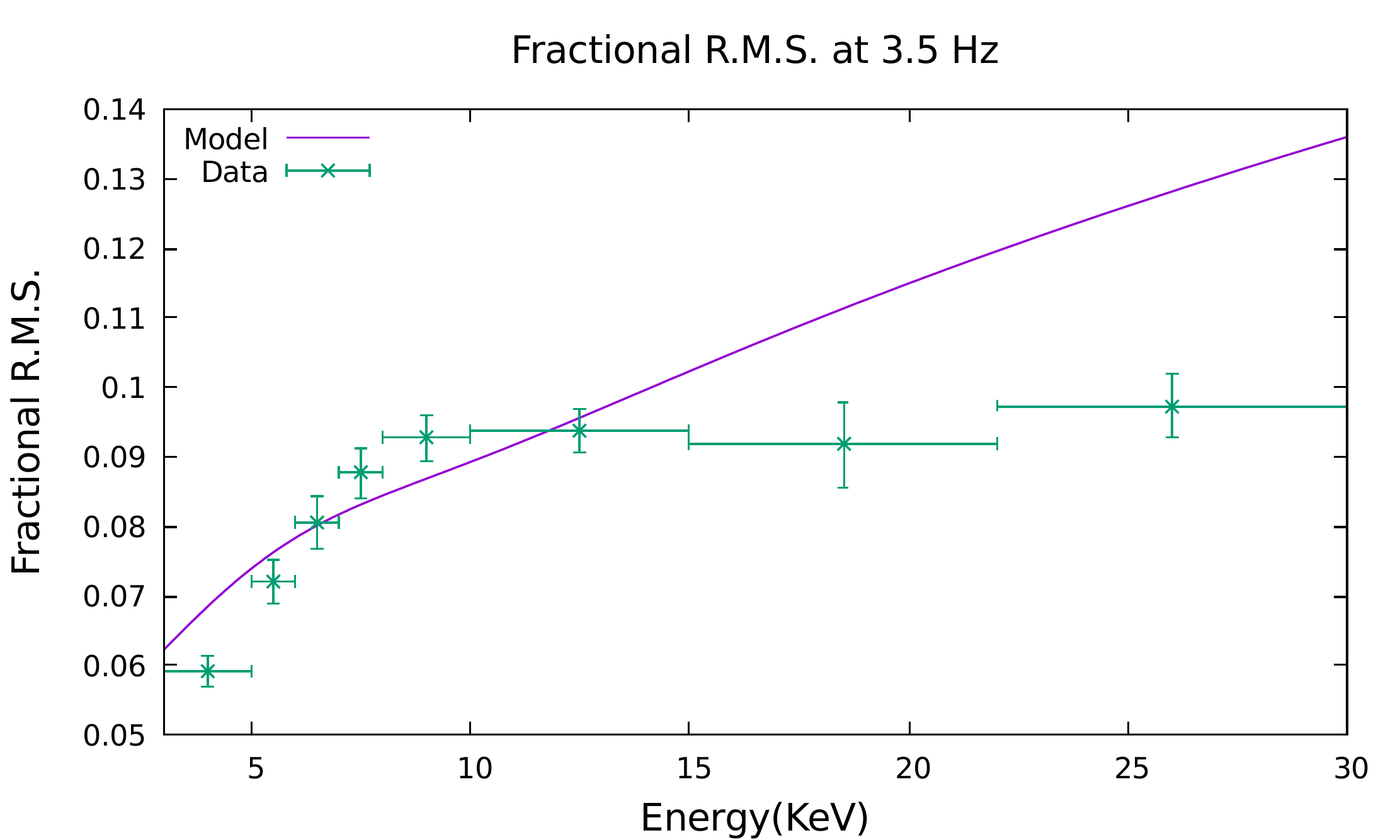}\hfill
\includegraphics[width=0.33\textwidth,height=4.8cm]{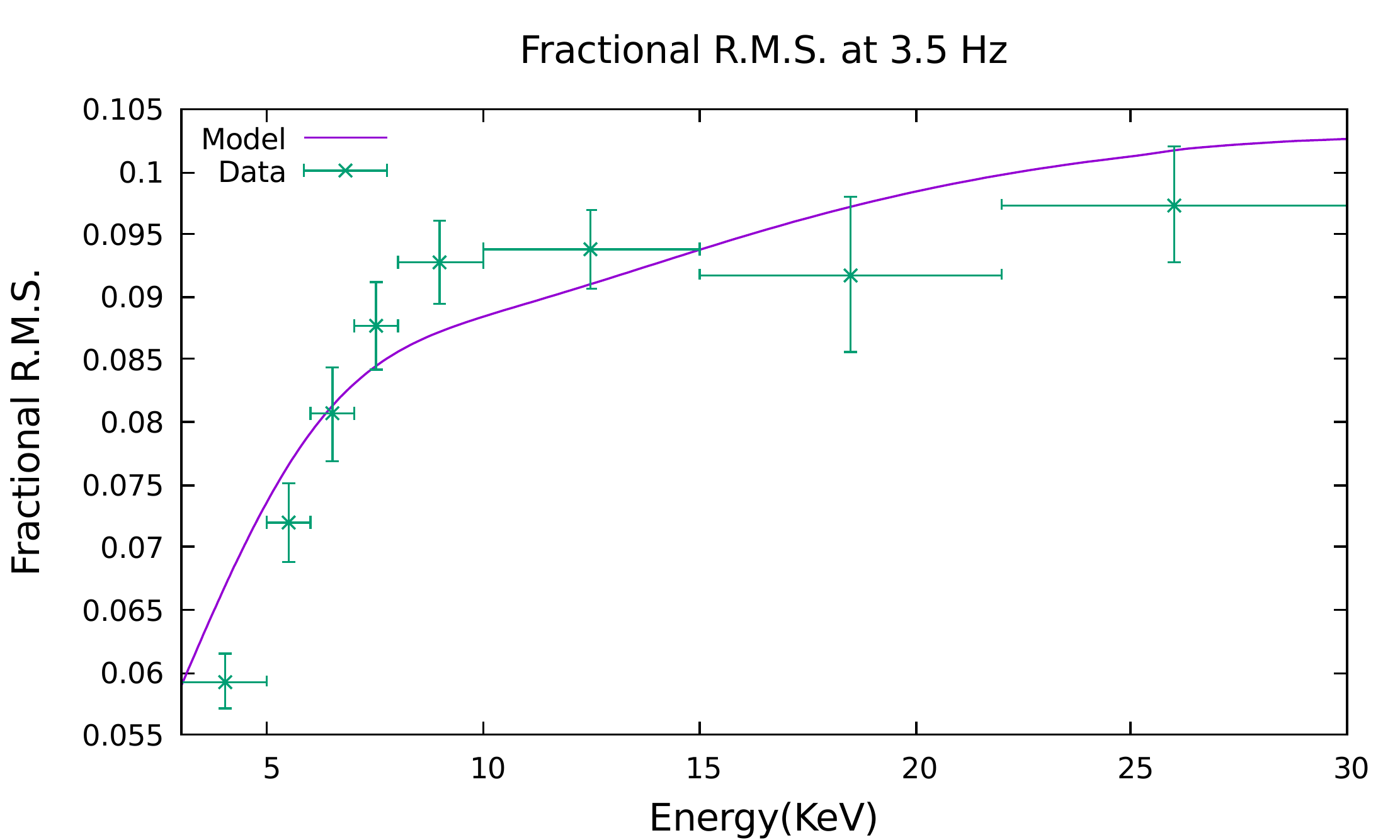}\hfill
\subfloat[]{\includegraphics[width=0.33\textwidth,height=4.8cm]{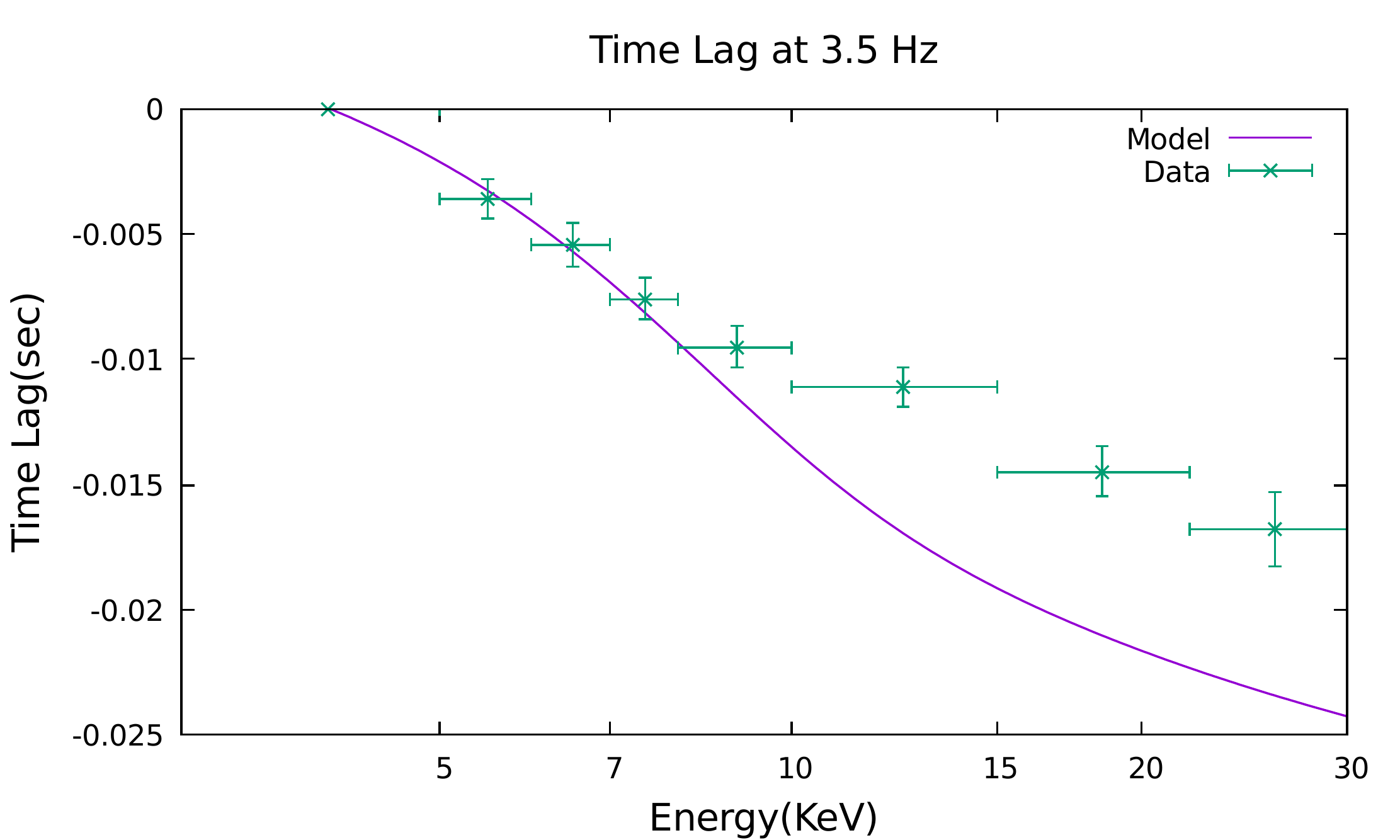}}\hfill
\subfloat[]{\includegraphics[width=0.33\textwidth,height=4.8cm]{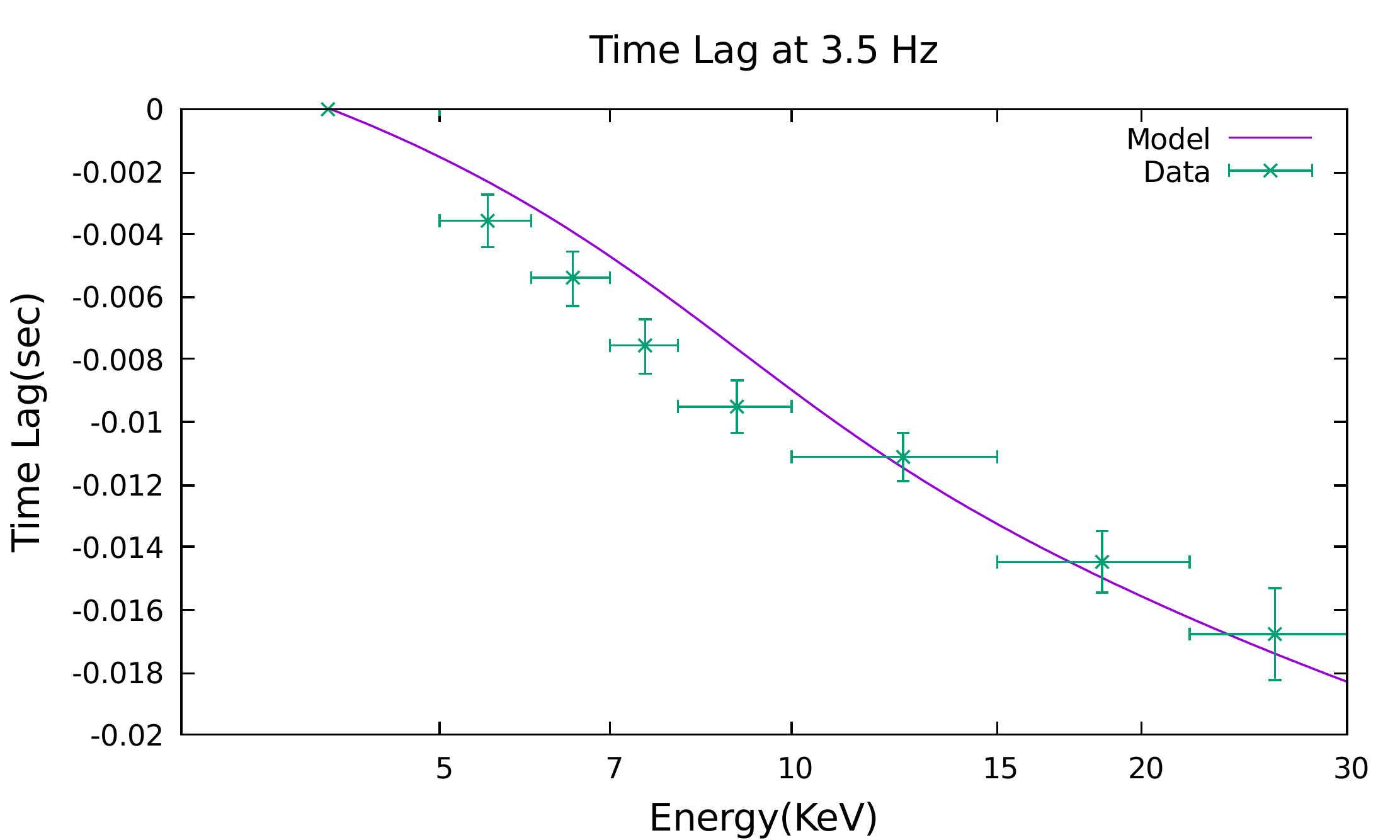}}\hfill
\subfloat[]{\includegraphics[width=0.33\textwidth,height=4.8cm]{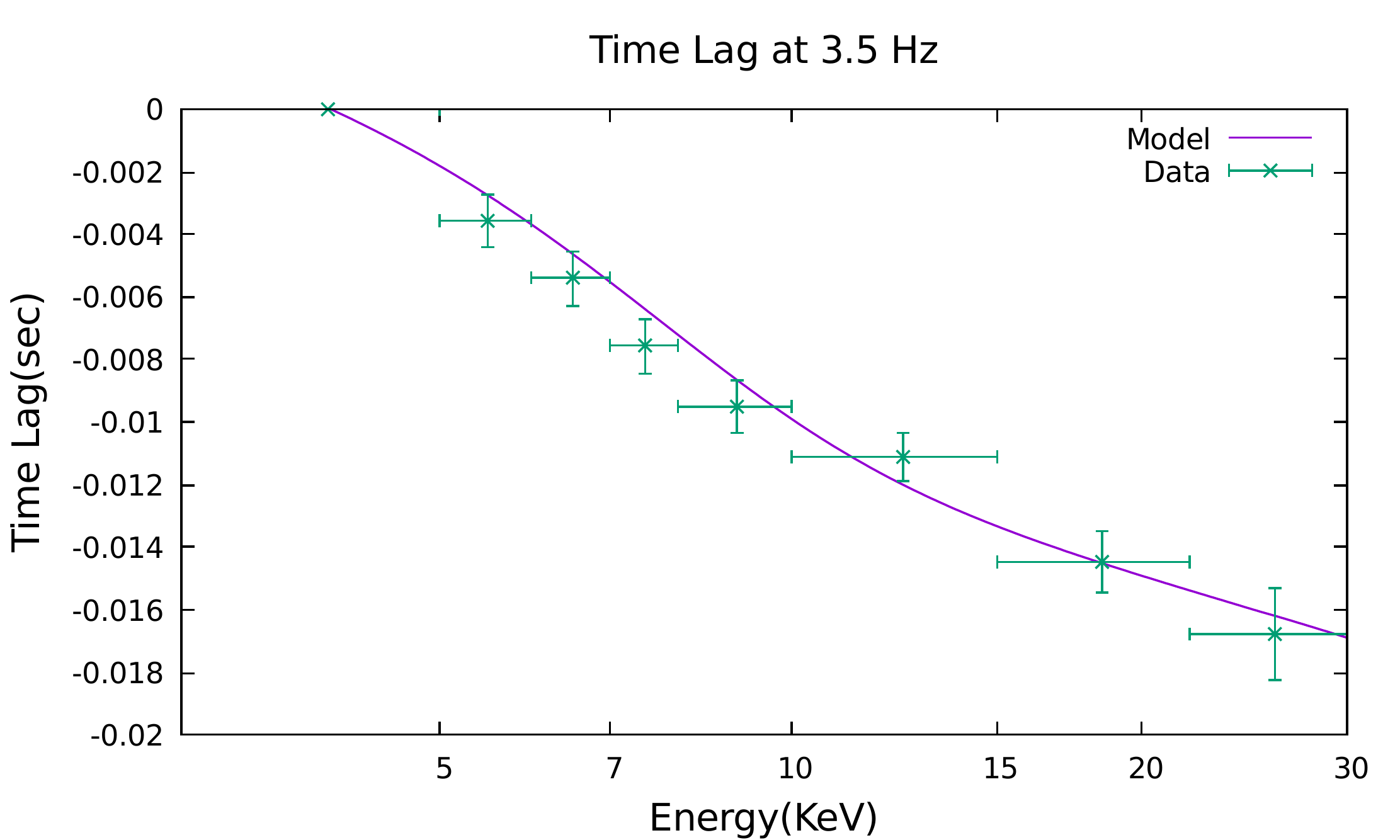}}

\caption{Top and bottom panels show observed Fractional rms and time lags versus energy at QPO frequency of 3.5 Hz for section I being fitted with  Model 1(Fig. a), Model 2(Fig.b) and Model 3(Fig.c) in the 3.0-30.0 KeV respectively.}
\label{fig:param2}
\end{figure*}

\begin{figure*}

\centering
\includegraphics[width=0.33\textwidth,height=4.8cm]{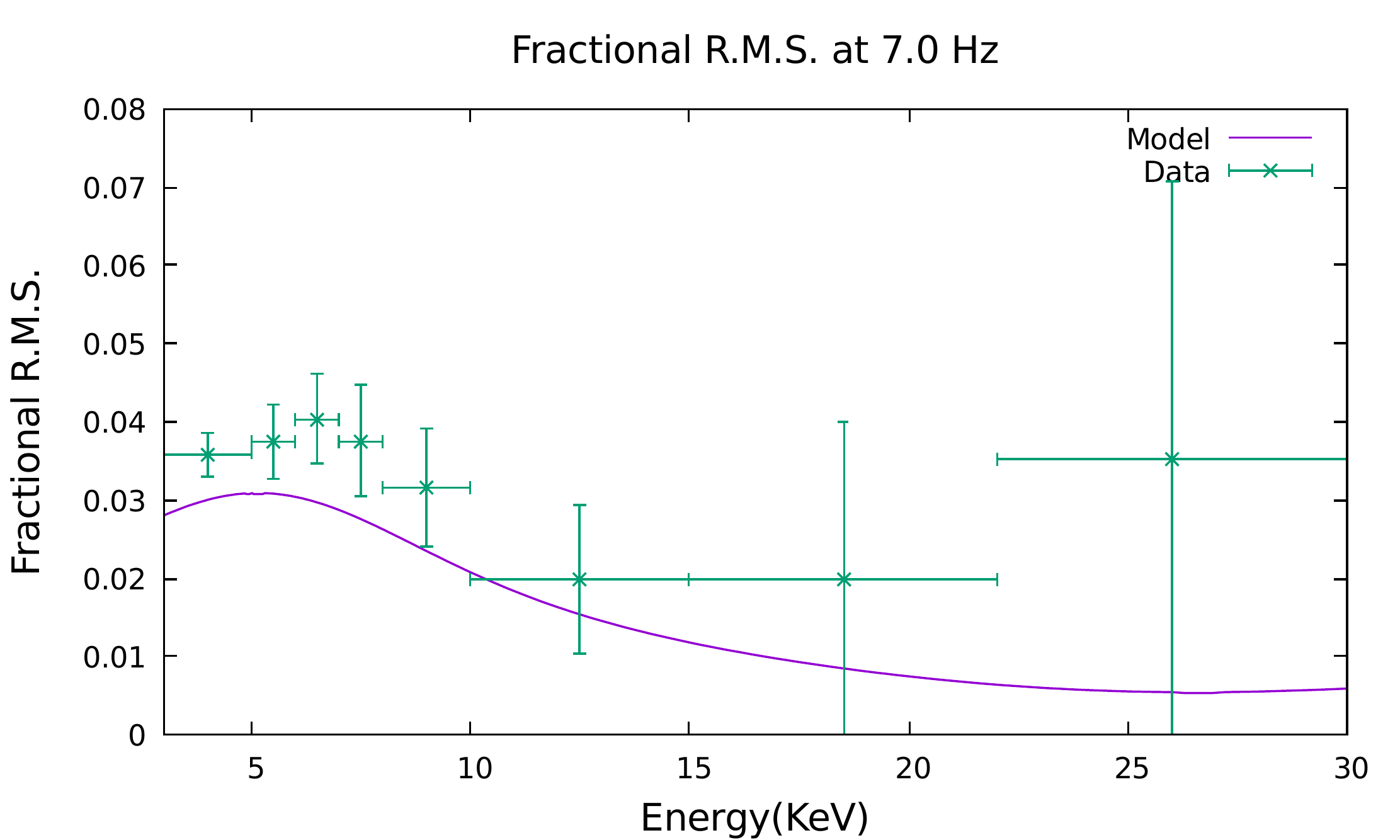}\hfill
\includegraphics[width=0.33\textwidth,height=4.8cm]{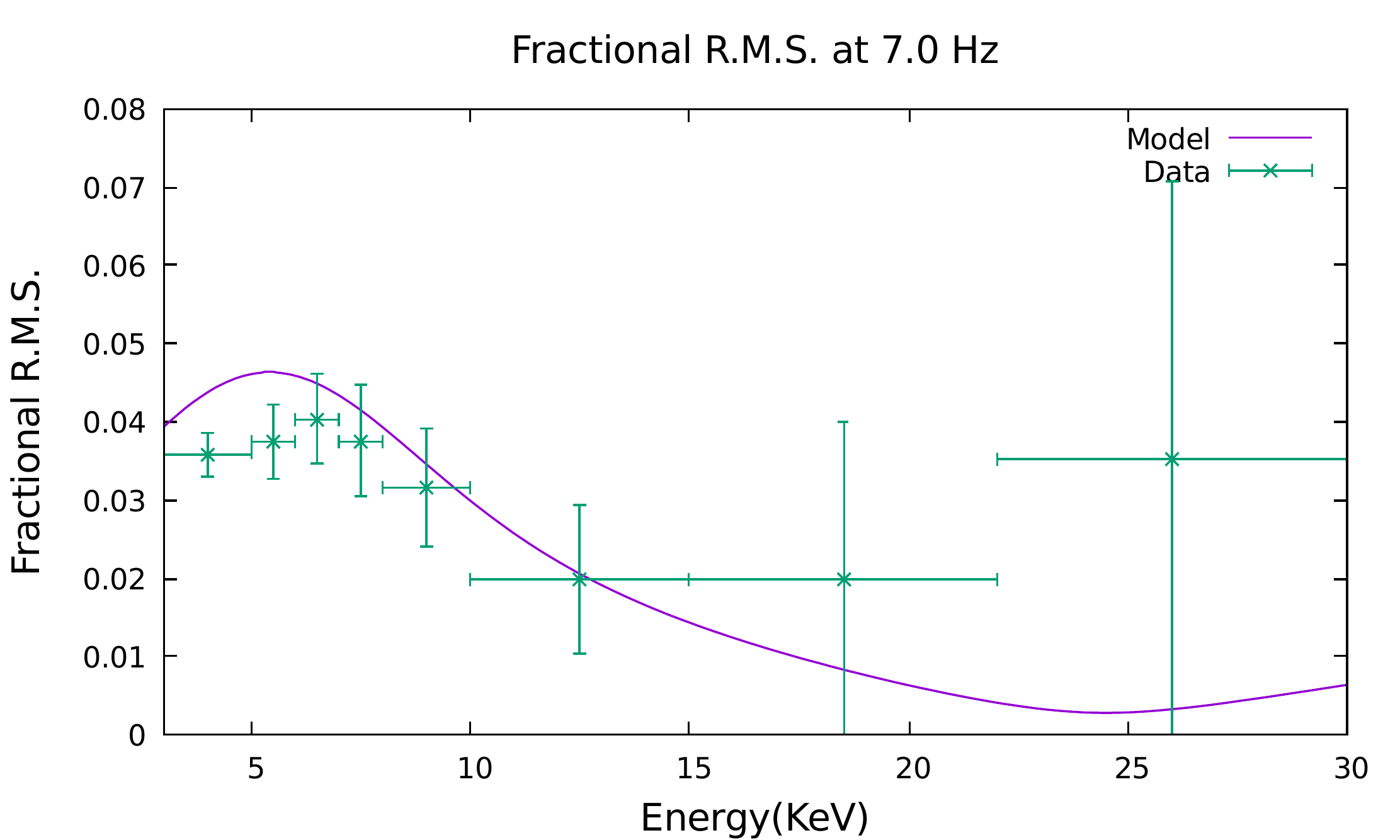}\hfill
\includegraphics[width=0.33\textwidth,height=4.8cm]{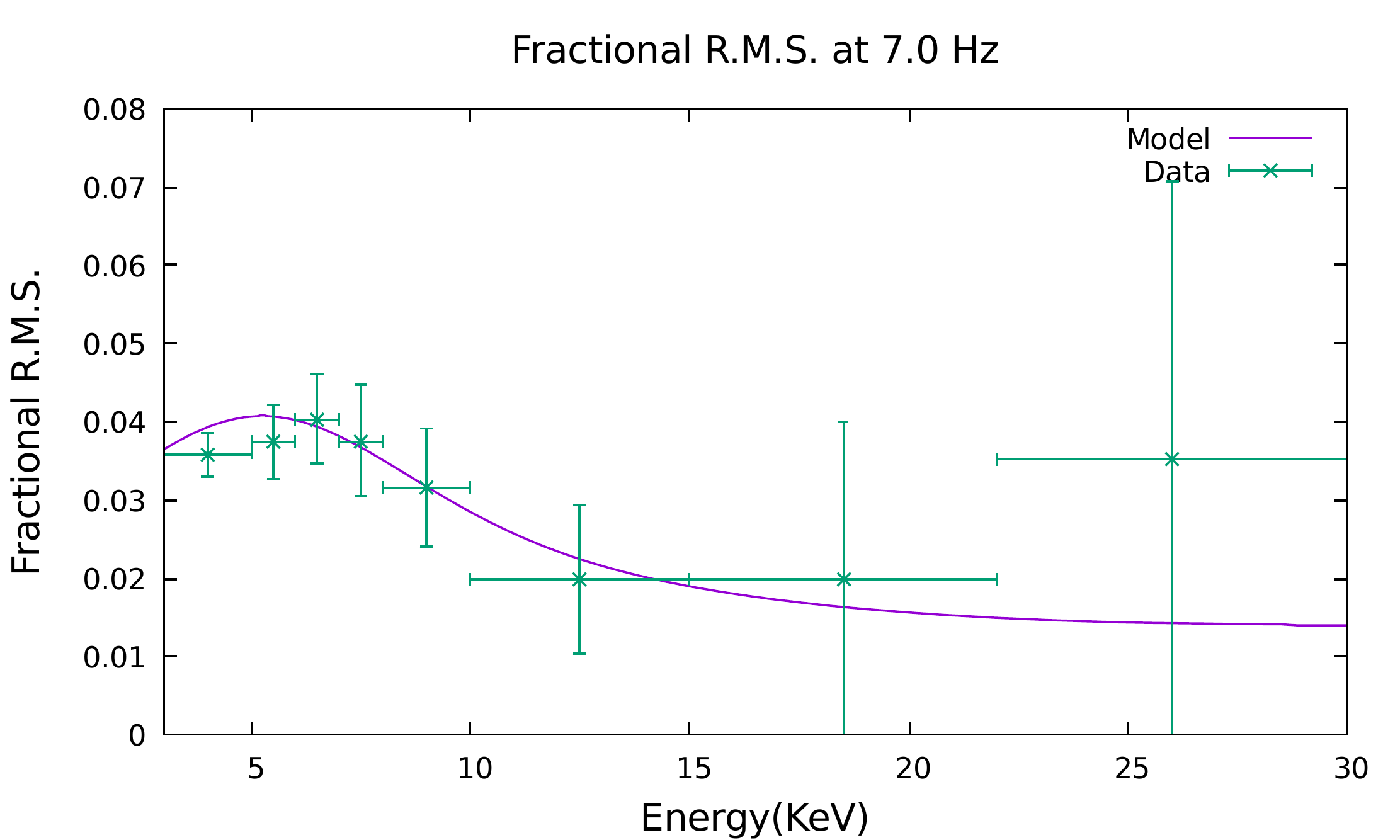}\hfill
\subfloat[]{\includegraphics[width=0.33\textwidth,height=4.8cm]{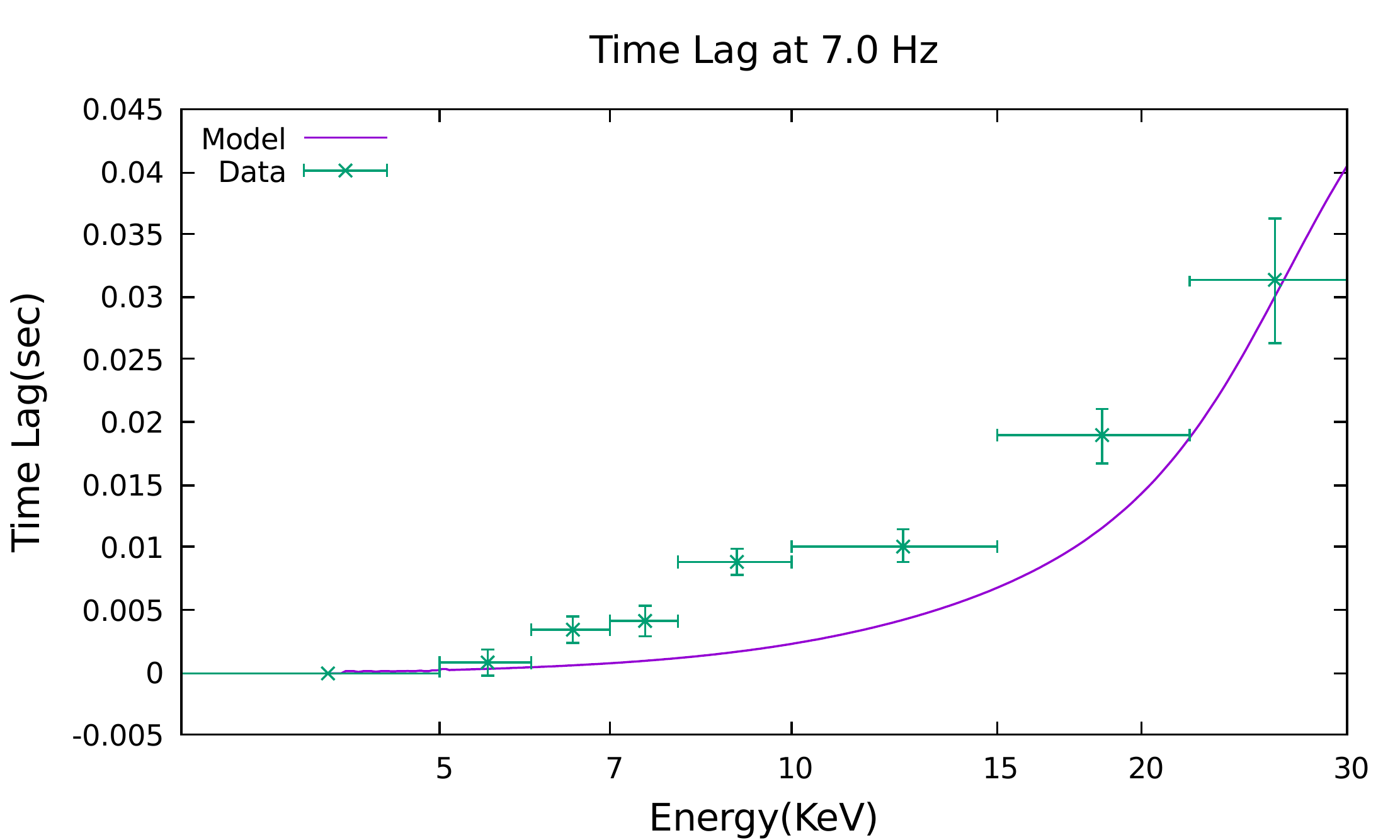}}\hfill
\subfloat[]{\includegraphics[width=0.33\textwidth,height=4.8cm]{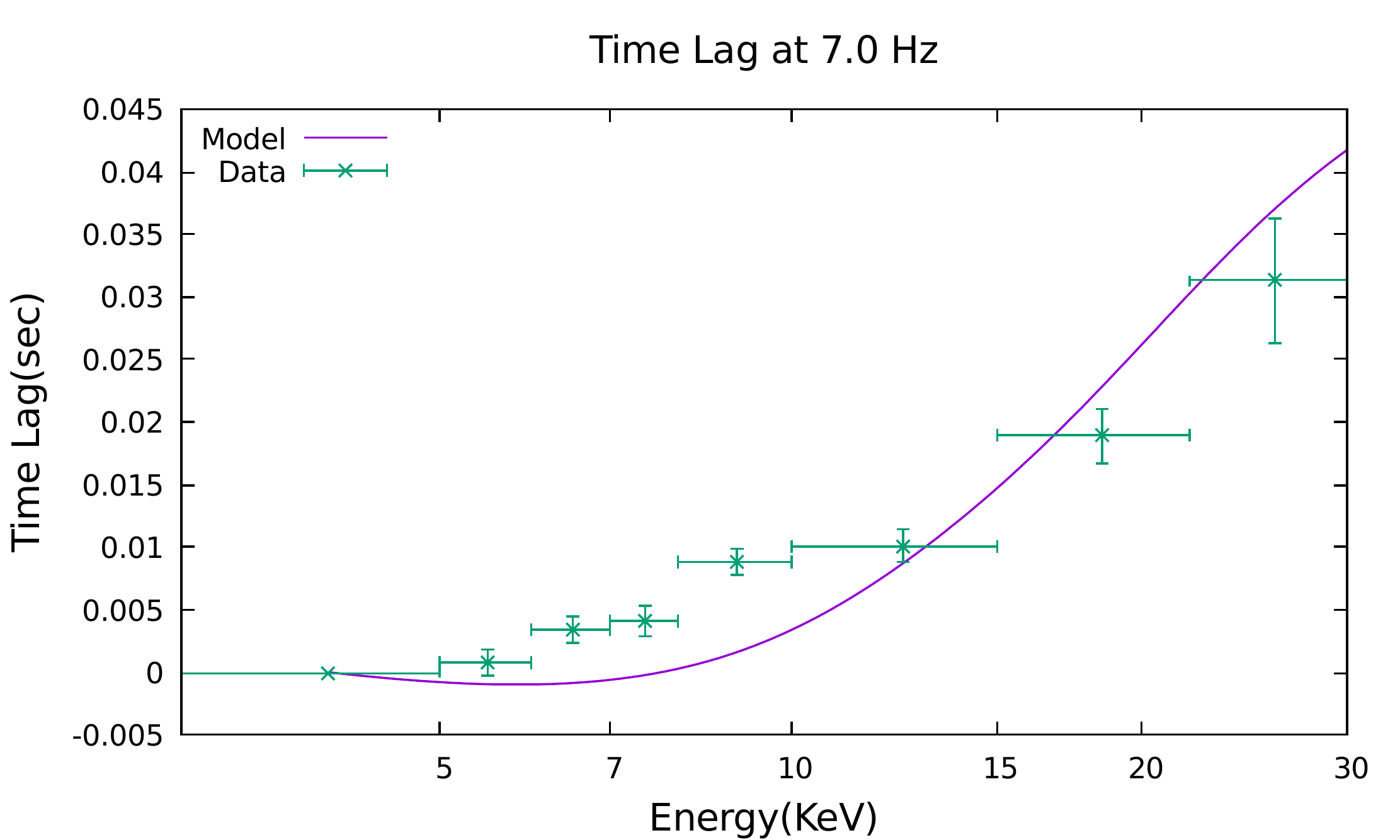}}\hfill
\subfloat[]{\includegraphics[width=0.33\textwidth,height=4.8cm]{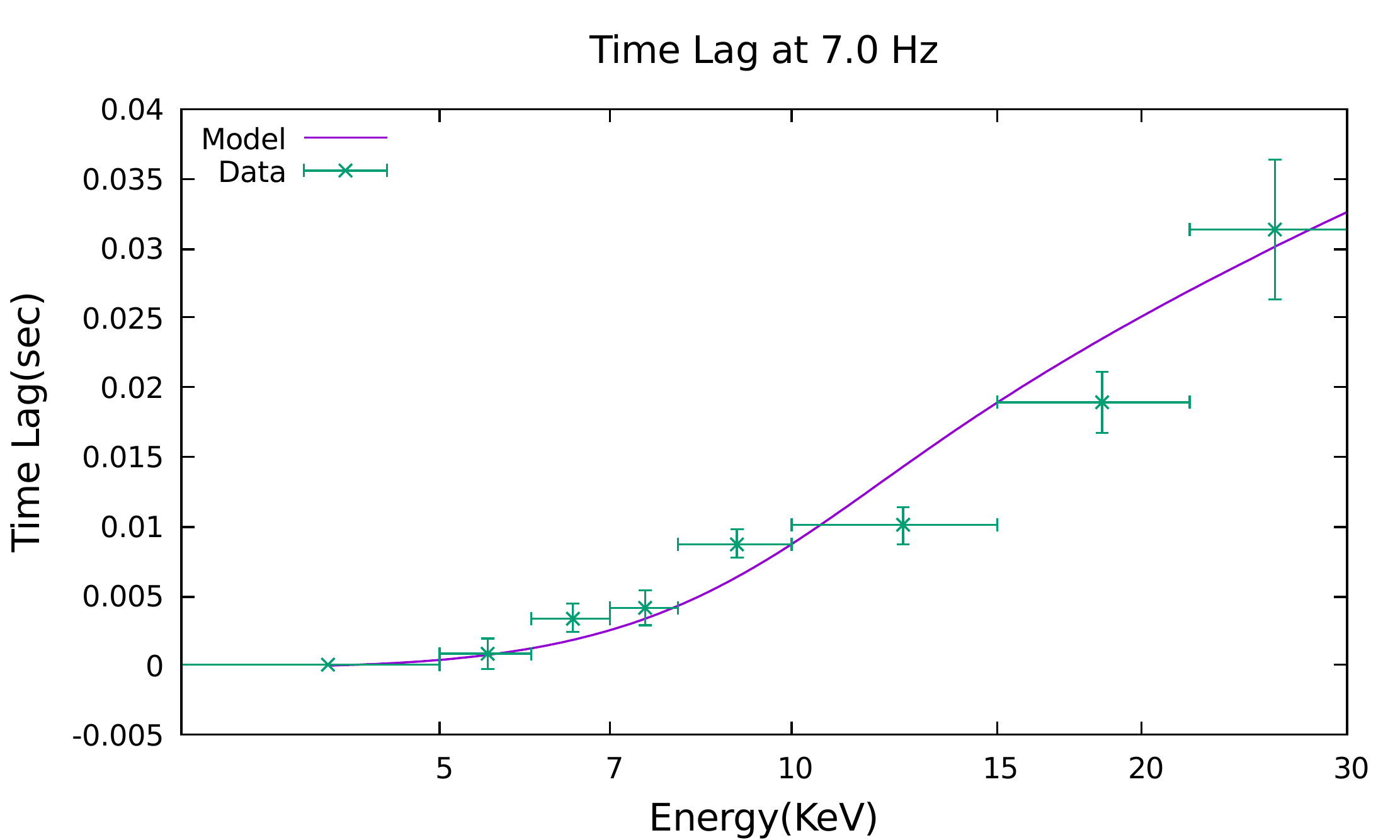}}

\caption{Top and bottom panels show observed Fractional rms and time lags versus energy at harmonic frequency of 7.0 Hz for section I  being fitted with predicted Model 1(Fig.a), Model 2(Fig.b) and Model 3(Fig.c) in the 3.0-30.0 KeV respectively.}
\label{fig:pmodel1}
\end{figure*}

To determine which physical parameter is responsible for the observed QPOs, we compare the predictions of the model described in the previous section with the energy dependent r.m.s and time-lag for the fundamental and harmonic frequencies for section I. We start with considering that only the inner disk temperature $kT_{in}$ and the heating rate $\dot H$ to vary with a time lag between them. This would correspond to the scenario where the coronal heating rate varies and then after a time delay there is a variation in the accretion rate of the disk resulting in variation of the inner disk temperature. We find that this simple model referred to as Model 1 cannot adequately describe the observed energy dependence of the r.m.s and time-lag for section I. This is illustrated in Figures \ref{fig:param2}a and \ref{fig:pmodel1}a, where the solid lines show the predictions of the model as compared with the observations for the QPO at $\sim 3.5$ Hz. Here the absolute values of the variations and the time-lag between them was varied to get as close a fit as possible, but there were significant deviations.\\

Next we attempted to describe the data by also varying the normalization of the disk component which would imply a variation in the Comptonized component normalization as well. Such a normalization variation represents a variation in the inner radius, along with the accretion rate. This model referred to as Model 2, has five parameters. Three parameters for the fractional values of the variations, $|\delta kT_{in}|$,$|\delta \dot H|$, $|\delta N_{dbb}|$ and two time-lags of $|\delta kT_{in}|$ and $|\delta N_{dbb}|$ with respect to the reference $|\delta \dot H|$. As can be seen in Figures \ref{fig:param2}b and \ref{fig:pmodel1}b, the model predictions are similar to those observed, although there are discrepancies especially for the energy dependent r.m.s of the fundamental QPO frequency. The physical picture which emerges for this model is that there is an coronal heating rate variation, followed  by a variation in the inner radius ($|\Delta N_{dbb}|$) and finally after a delay there is a change in the accretion rate(giving rise to $|\Delta KT_{in}|$) as the perturbation moves outwards.\\

\begin{figure*}

\centering
\includegraphics[width=0.5\textwidth,height=5cm]{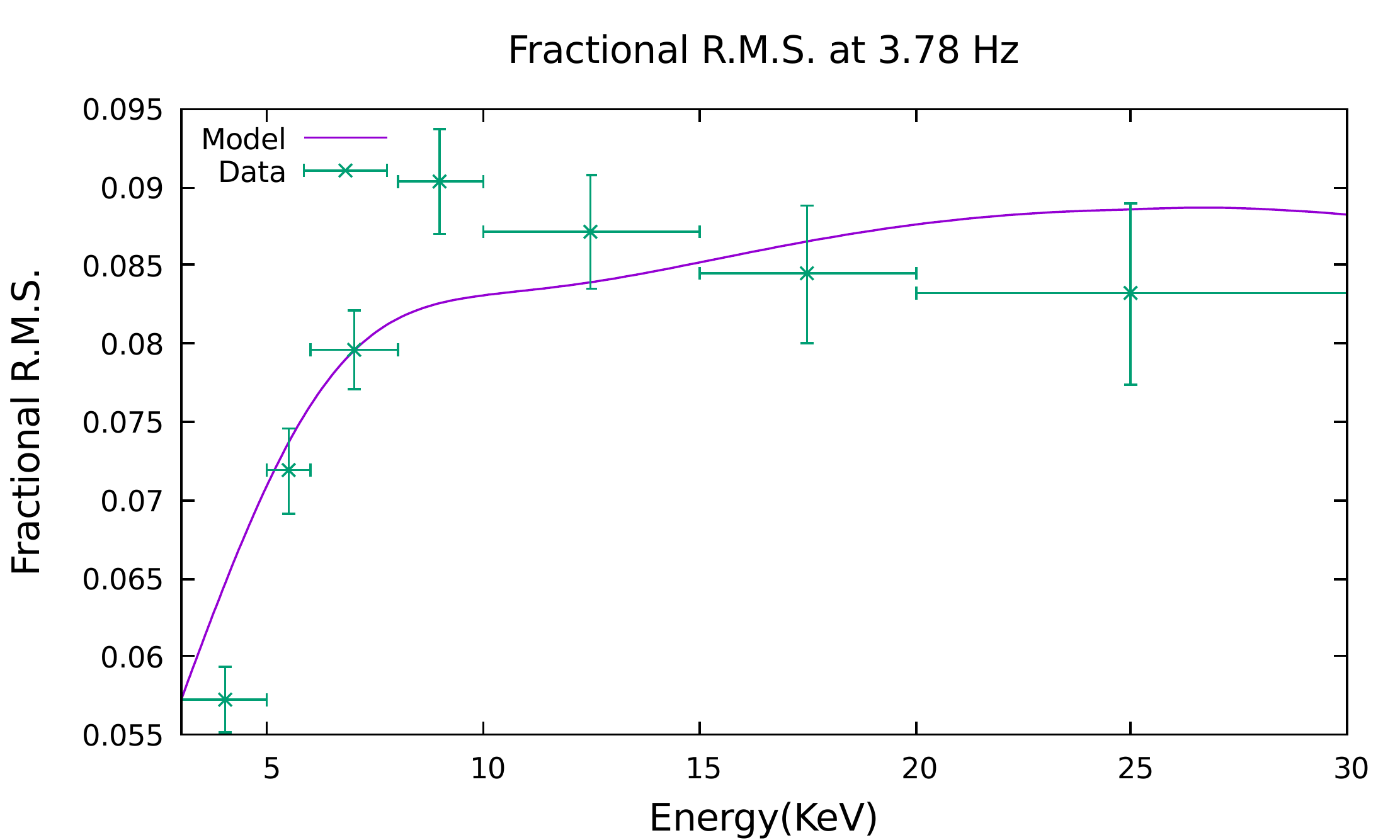}\hfill
\includegraphics[width=0.5\textwidth,height=5cm]{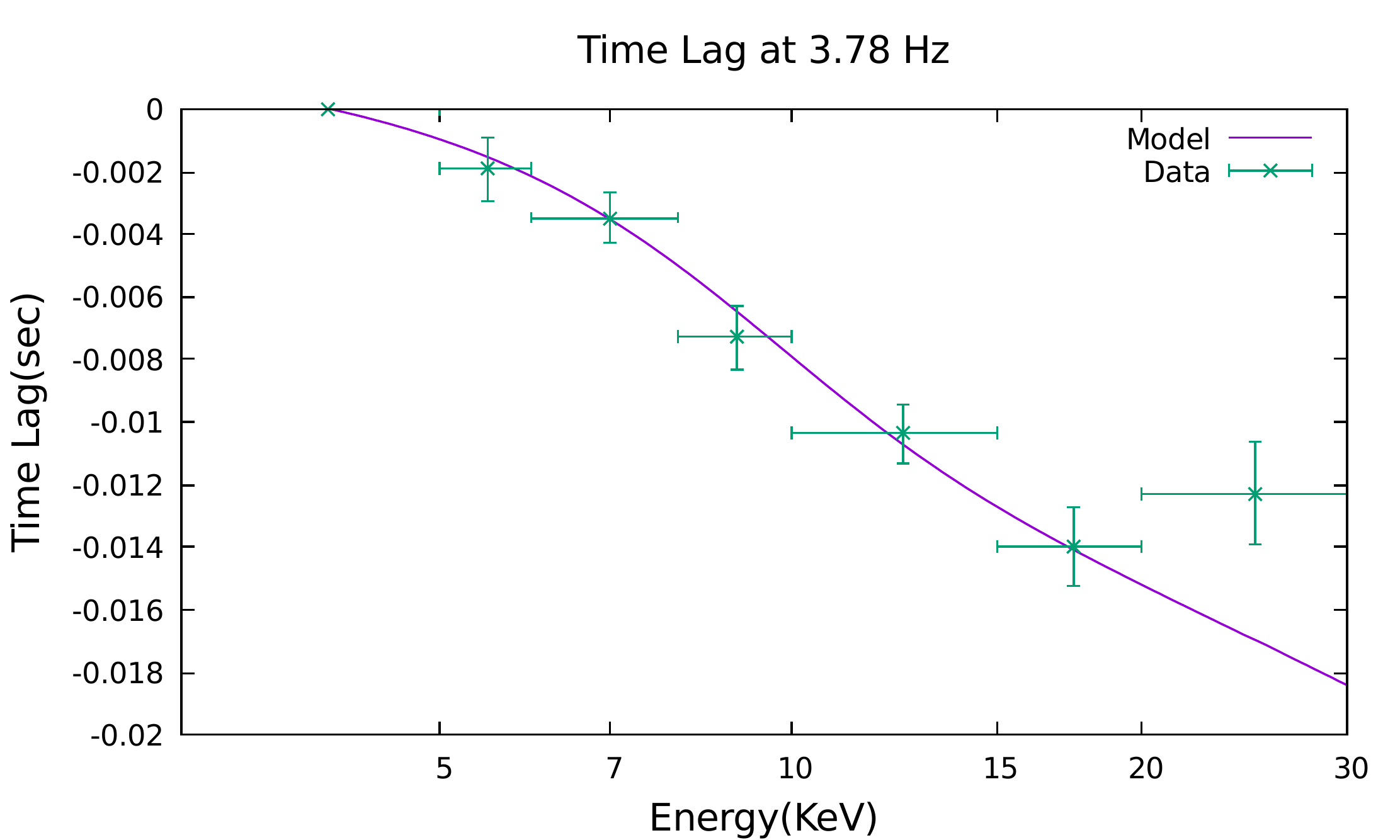}\hfill
\includegraphics[width=0.5\textwidth,height=5cm]{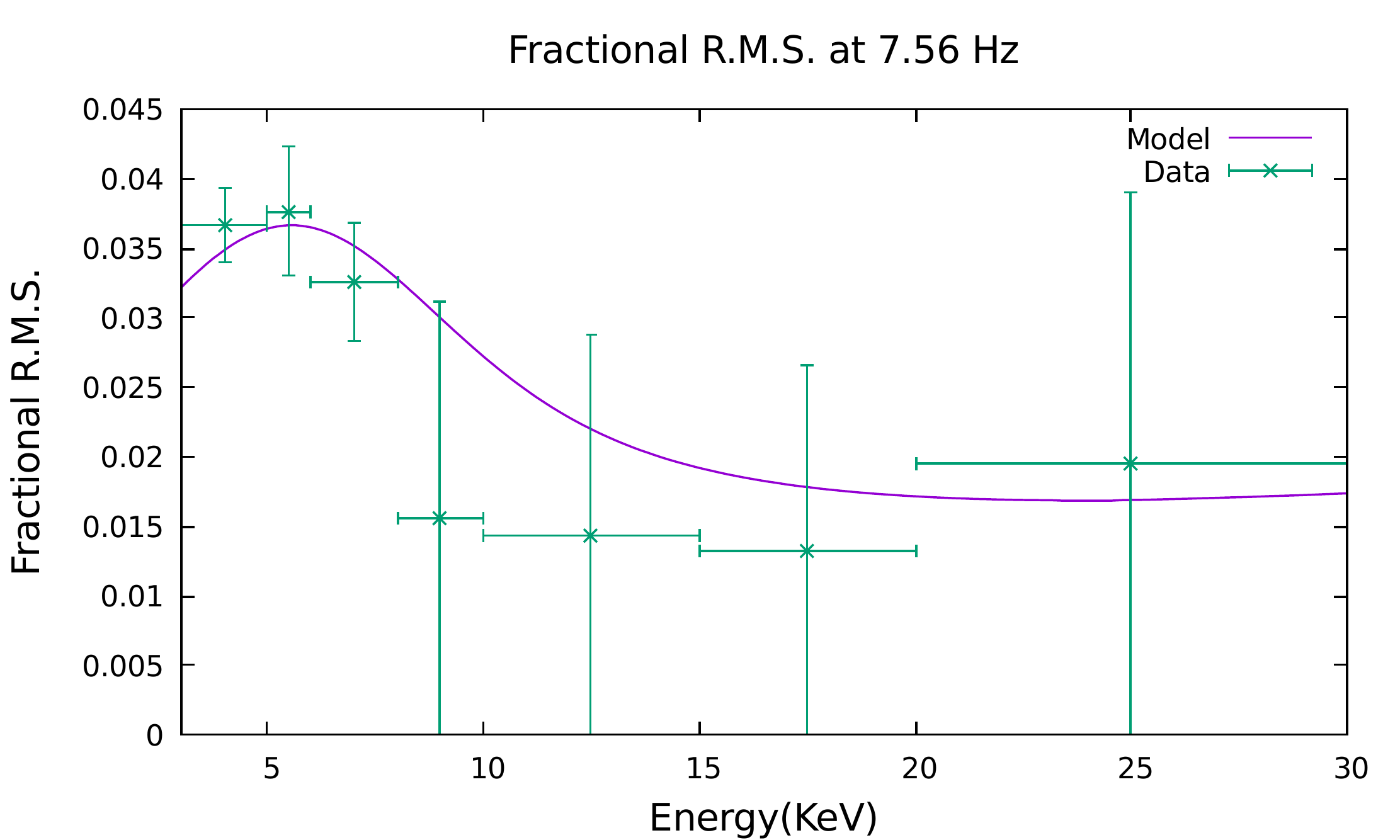}\hfill
\includegraphics[width=0.5\textwidth,height=5cm]{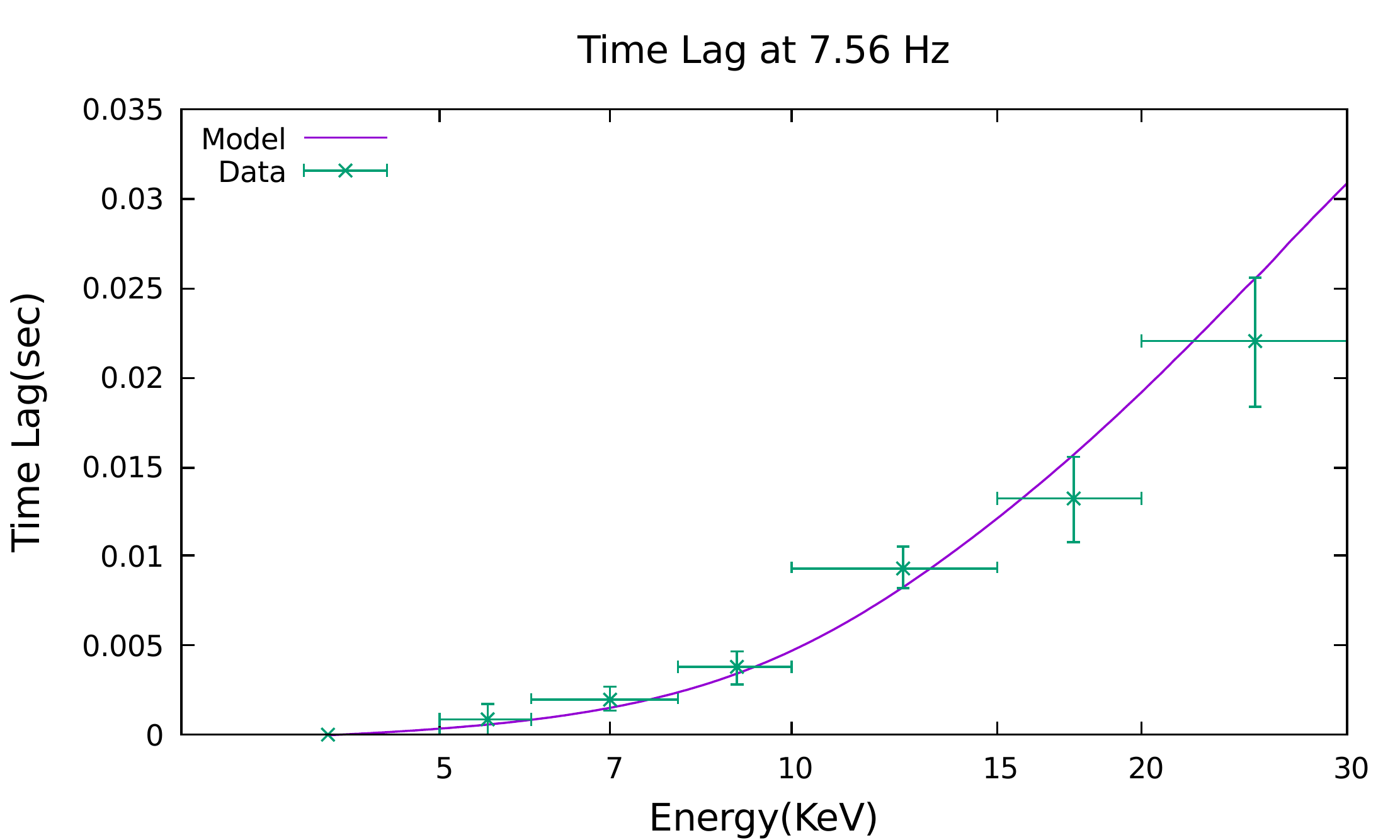}

\caption{Top and bottom panels show observed Fractional rms and time lags versus energy at QPO's fundamental frequency of 3.78 Hz and its harmonic for section II being fitted with Model 3 of seven parameters in the 3.0-30.0 KeV respectively.}
\label{fig:pmodel2}
\end{figure*}

\begin{figure*}

\centering
\includegraphics[width=0.5\textwidth,height=5cm]{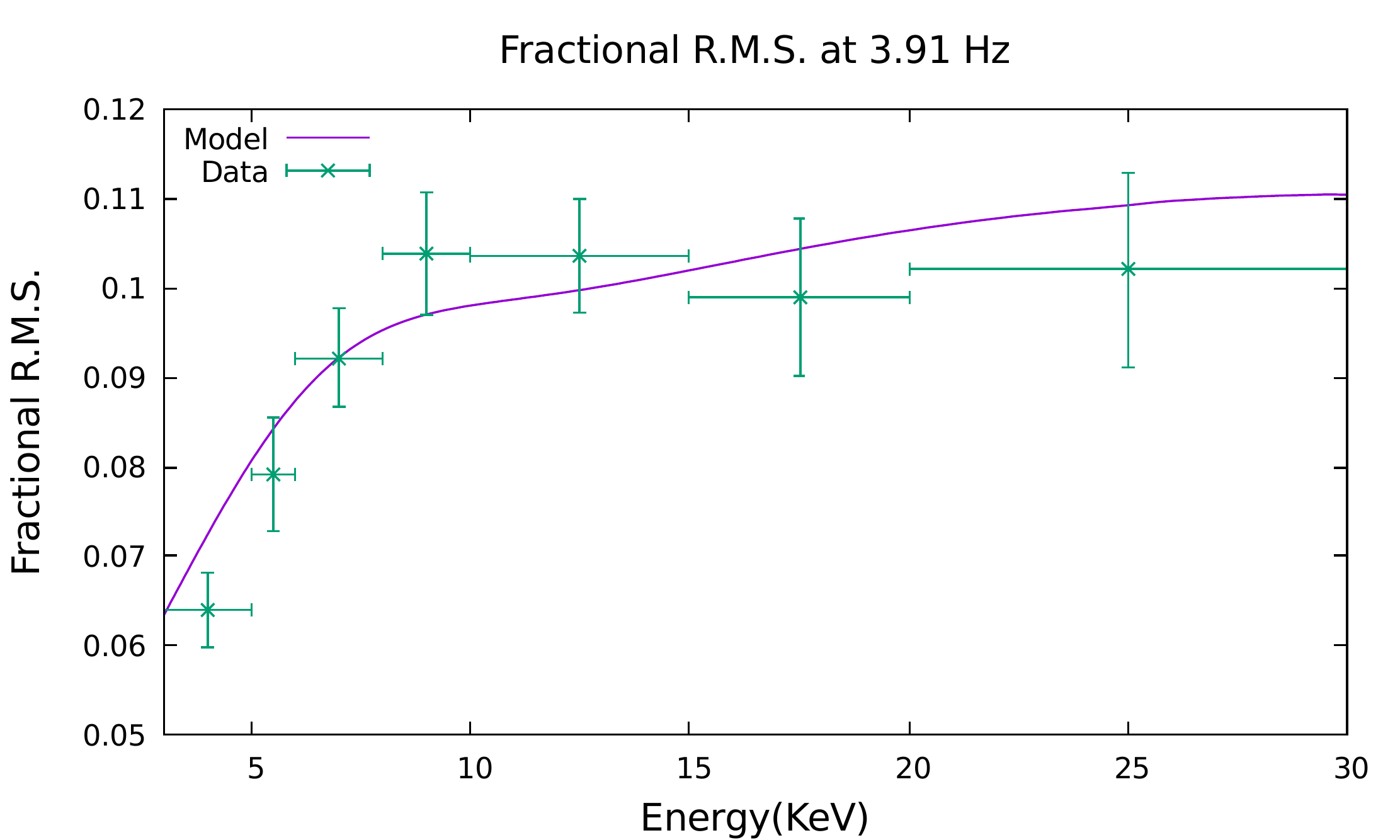}\hfill
\includegraphics[width=0.5\textwidth,height=5cm]{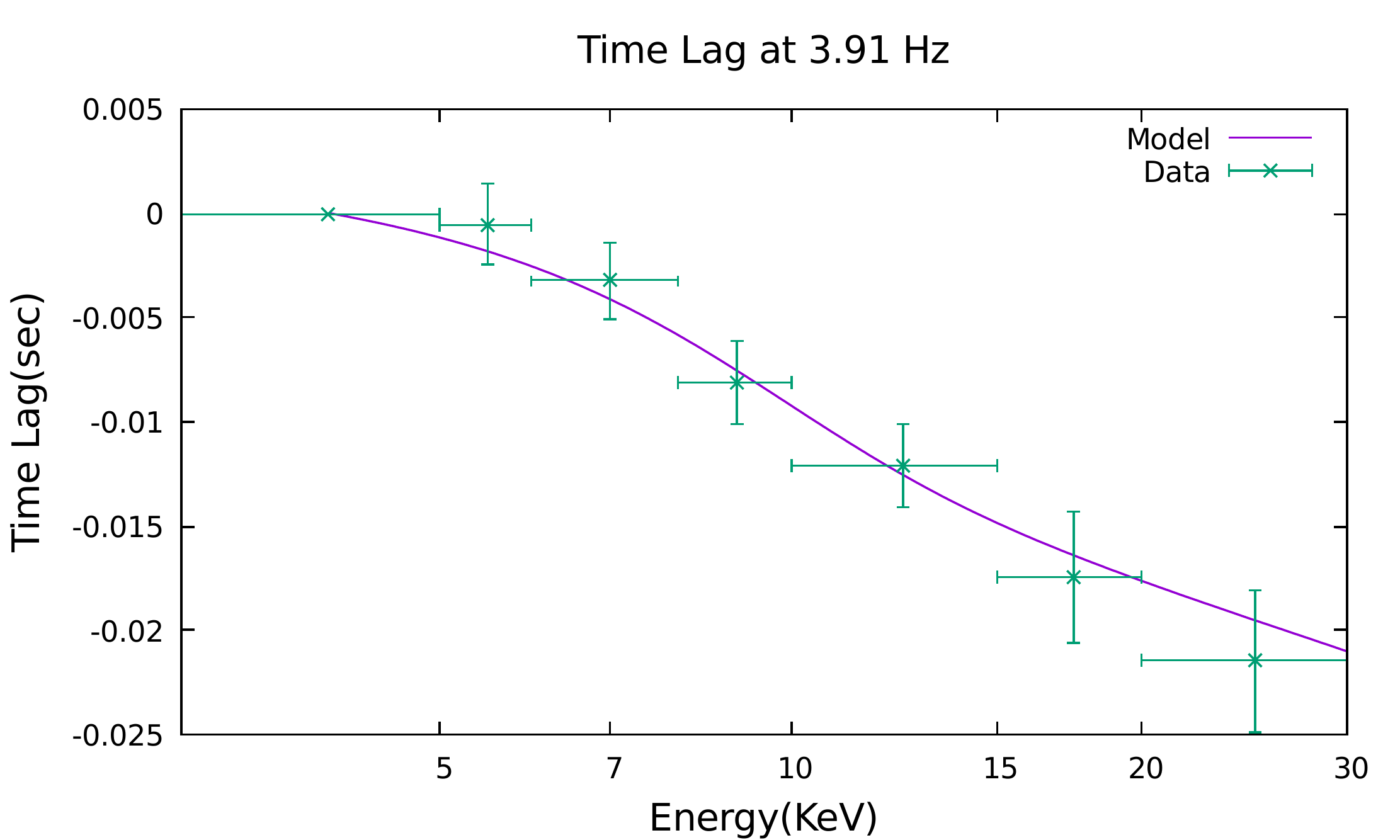}\hfill
\includegraphics[width=0.5\textwidth,height=5cm]{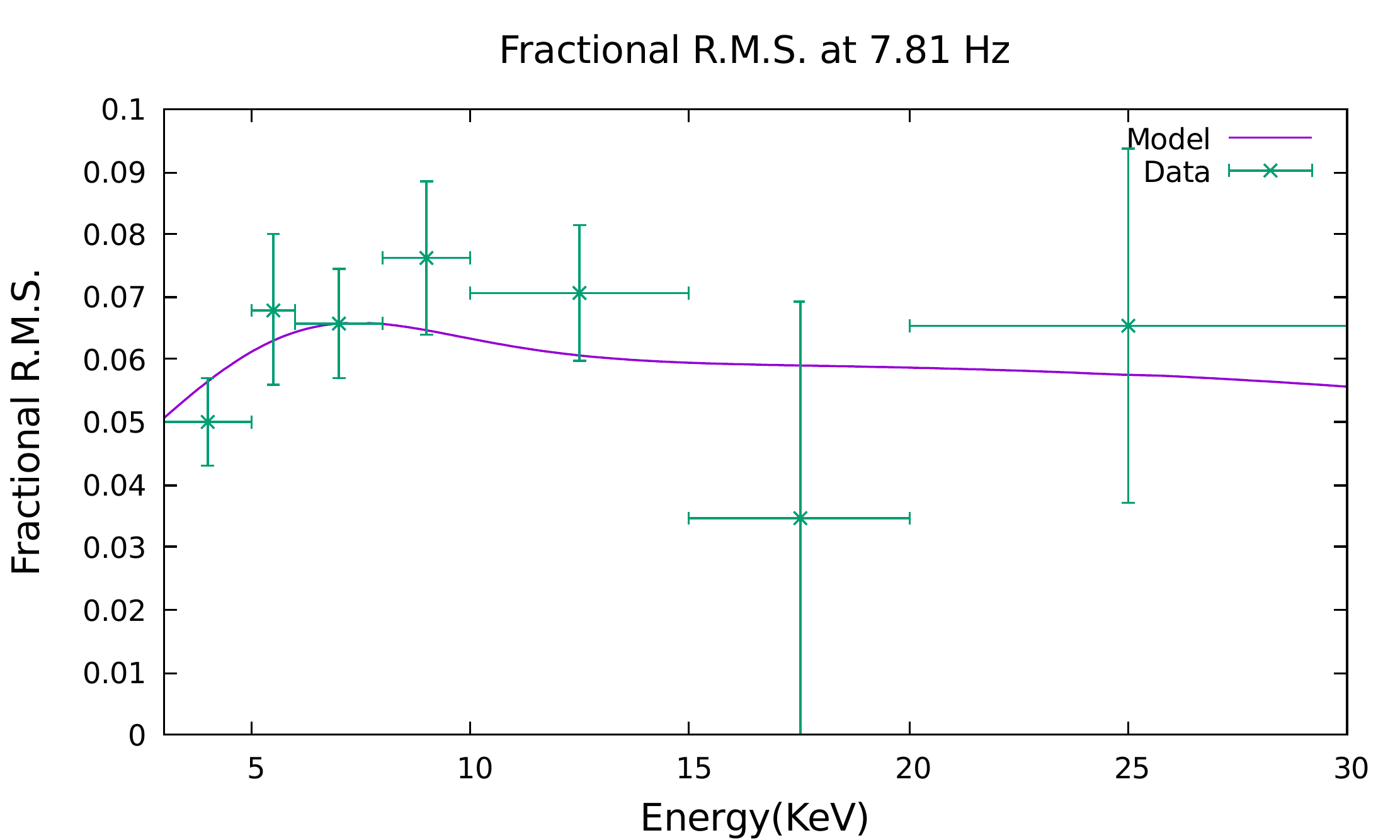}\hfill
\includegraphics[width=0.5\textwidth,height=5cm]{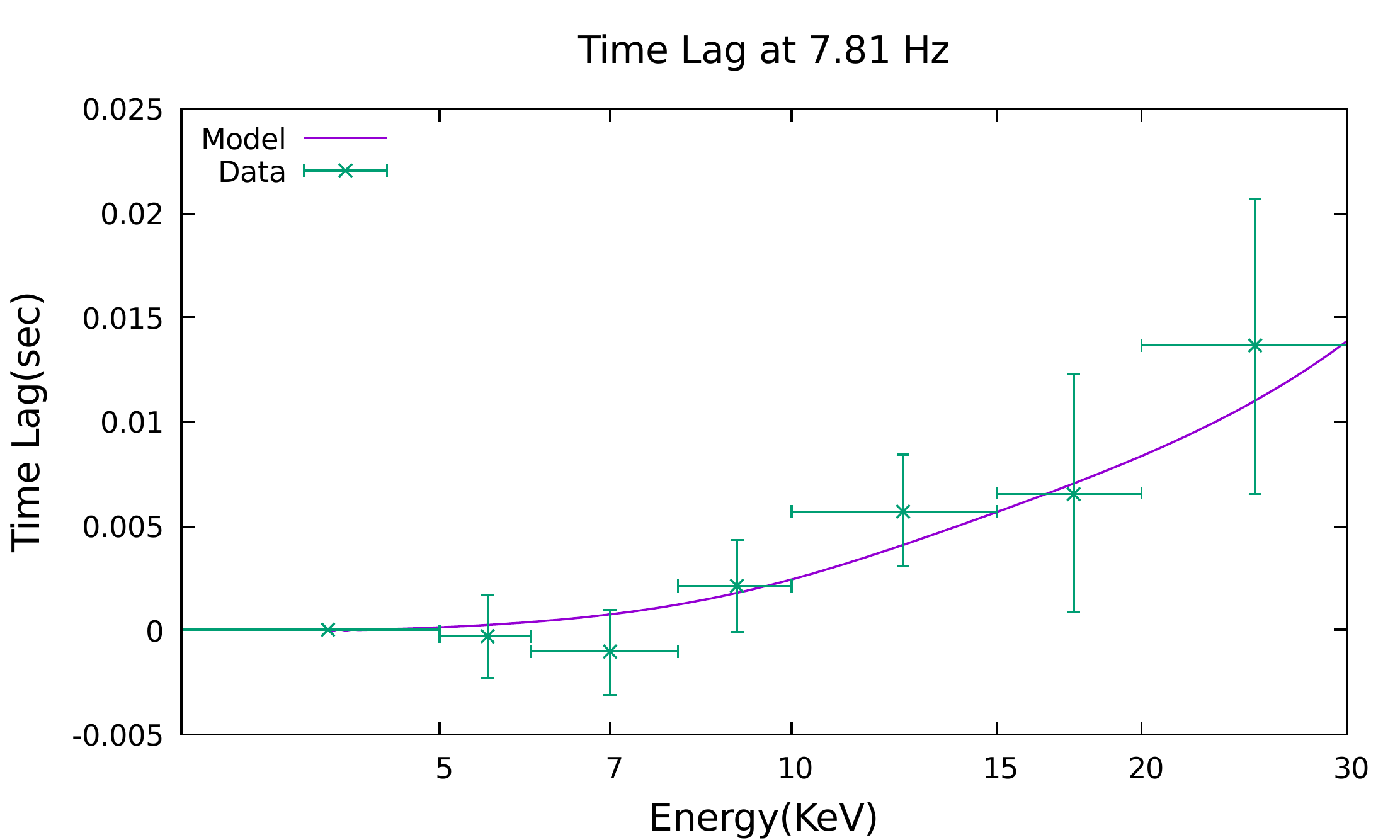}

\caption{Top and bottom panels show observed Fractional rms and time lags versus energy at QPO's fundamental frequency of 3.91 Hz and its harmonic for section III being fitted with predicted Model 3 of seven parameters in the 3.0-30.0 KeV respectively.}
\label{fig:pmodel3}
\end{figure*}

\begingroup
\setlength{\tabcolsep}{4pt} 
\renewcommand{\arraystretch}{1.5}
 \begin{table*}
    \centering
     \caption{Model Parameters for three sections of data}
    \begin{tabular}{c c c c c c c c c c c}
    \hline \hline
		 & & & & & Sections & & & & &\\
		\hline    
         Parameters & Description/Unit & \multicolumn{2}{c|}{I} & \multicolumn{2}{c|}{II} & & & \multicolumn{2}{c|}{III} \\
         \hline
          &  &  Funda. & Harmo. & Funda. & \multicolumn{2}{c|}{Harmo.} & \multicolumn{2}{c|}{Funda.} & \multicolumn{2}{c|}{Harmo.}\\
          \hline
          & & M3 & M3 & M3 & M3 & M1 & M3 & M1 & M3 & M1 \\
          \hline
         
          $\tau_1$& Time lag(msec) of   &  $22.7^{+1.9}_{-2.2}$ & $104.6^{+1.2}_{-1.2}$ & $23.5^{+1.2}_{-1.2}$ & $103.2^{+7.3}_{-9.5}$ & $103.2^{+2.9}_{-2.9}$ &  $24.4^{+5.4}_{-6.5}$ & $24.4^{+4.7}_{-4.7}$ & $116.2^{+24.8}_{-17.6}$ & $116.2^{+8.6}_{-8.6}$\\   
                & $\delta (kT_{in})$ w.r.t. $\delta \dot{H}$\\

          $\tau_2$& Time lag(msec) of   &  $9.0^{+1.5}_{-6.8}$ & $6.8^{+1.1}_{-1.3}$ & $4.2^{+1.6}_{-1.2}$ & $<9.8$ & - & $<9.3$ & - & $<29.1$ & - \\ 
          & $\delta (\tau)$ w.r.t. $\delta \dot{H}$\\
          
          $\tau_3$& Time lag(msec) of  &  $13.6^{+0.5}_{-1.8}$ & $<12.5$ & $8.4^{+4.5}_{-4.8}$ & $<11.5$ & - &  $12.2^{+2.7}_{-5.5}$ & - & $<28.2$ & - \\
          & $\delta (N_{dbb})$ w.r.t. $\delta \dot{H}$\\
          
          $\mid \delta (kT_{in})\mid$& Variation in inner  &  $1.90^{+0.01}_{-0.01}$ & $1.40^{+0.03}_{-0.07}$ & $1.80^{+0.01}_{-0.01}$ & $1.15^{+0.05}_{-0.05}$ & $1.15^{+0.02}_{-0.13}$ & $2.16^{+0.02}_{-0.04}$ & $2.24^{+0.04}_{-0.15}$ & $3.50^{+0.13}_{-0.13}$ &  $2.2^{+0.02}_{-0.02}$\\
          & disk temperature(\%)\\
          
          $\mid \delta \dot{H} \mid$& Variation in      &  $11.0^{+0.5}_{-0.3}$ & $1.0^{+0.1}_{-0.2}$ & $5.2^{+0.4}_{-0.4}$ & $1.0^{+0.5}_{-0.1}$ &  $1.90^{+0.04}_{-0.04}$ & $9.0^{+0.7}_{-1.2}$ &  $16.0^{+2.2}_{-0.4}$ & $1.8^{+4.6}_{-0.9}$ & $7.0^{+3.4}_{-1.1}$ \\
          & heating rate(\%)\\
          
          $\mid \delta \tau \mid$& Variation in         &  $>87.0$  & $30.0^{+1.3}_{-2.0}$  & $>97.0$  & $15.0^{+4.4}_{-1.9}$  & - & $>86.0$ & -  & $>61.0$ & -  \\
          & optical depth(\%)\\
          
          $\mid \delta N_{dbb} \mid$& Variation in     &  $0.10^{+0.03}_{-0.03}$ & $0.90^{+0.07}_{-0.06}$ & $0.70^{+0.05}_{-0.07}$ & $0.1^{+0.4}_{-0.5}$ & - & $0.1^{+0.1}_{-0.1}$ & - & $0.9^{+0.9}_{-0.9}$ & - \\
          & disk Normalisation(\%)\\
          
          \hline
          $\chi^2$/dof & Reduced chi-square           &  26.31/8  & 24.95/8  & 28.82/6  & 6.01/6  & 13.05/10 & 7.79/6 & 12.43/10 & 4.72/6 & 12.60/10 \\
        \hline    
    \end{tabular}
    \label{tab:model}
    Note: M1 and M3 refer to the Model 1 and Model 3 respectively.
\end{table*} 
\endgroup
Further, in order to improve upon the Model 2 predictions, we consider variation in the optical depth as well. Such a model(Model 3) has seven parameters now- four for the $|\delta kT_{in}|$,$|\delta \dot H|$, $|\delta N_{dbb}|$, $|\delta \tau|$  and three time-lags of $|\delta kT_{in}|$, $|\delta N_{dbb}|$ and $|\delta \tau|$  with respect to the reference  $|\delta \dot H|$. Figures \ref{fig:param2}c, \ref{fig:pmodel1}c, \ref{fig:pmodel2} and \ref{fig:pmodel3} show that the Model 3 predictions quantitatively fit the observed temporal behavior for the three sections. The resulting interpretation in this case will be then that the change in coronal heating rate precedes the variations in the optical depth and thereby driving the oscillations in inner radius resulting in the accretion rate variations as perturbation moves outwards.\\

To quantify the improvement of the fitting for Model 3, we computed the reduced $\chi^2$ for all the three models. For the primary QPO at 3.5 Hz for Section I, using Models 1, 2 and 3, the $\chi^2$/(degrees of freedom) for the joint r.m.s and time-lag fits
are 234.81/12, 97.52/10 and 26.31/8 respectively.  Similarly, for the corresponding harmonic at $\sim 7$ Hz, the reduced $\chi^2$ decreases from 105.31/12 in Model 1 and 96.21/10 in Model 2 to 24.95/8 in Model 3. The parameters values for Model 3 with 1-$\sigma$ errors are listed in Table \ref{tab:model}. Although there is qualitative improvement in the fitting for the more complex Model 3 as compared to Model 1 for data Section 1,
the reduced $\chi^2$ is still large and the fit is not acceptable statistically. We note however, that the motivation here is to make a
qualitative comparison with predictions of the model described in this paper with data. A rigorous statistical analysis should take into
account several other factors such as the errors in the time averaged spectral parameters while fitting the r.m.s and time-lag. Moreover,
here we have just compared the data value with the theoretical one expected at the mid point of the energy bin, while actually one
should integrate the model prediction within each energy bin, which should incorporate the variation of the effective area of the
instruments within that energy bin.\\

For Section II, for the fundamental QPO at 3.78 Hz, we again find that Model 3 provides a better description of the data
than Models 1 and 2 whereas for the harmonic at 7.56 Hz of Section II and for both QPOs of Section III, Model 1 adequately describes the
data with reduced $\chi^2 \sim 1$. Although we have listed(in Table \ref{tab:model}) the parameter values for Model 3 with 1-$\sigma$ errors for both QPOs of section I and fundamental QPO of section II, for the harmonic of section II and both QPOs of section III we have enlisted the parameter values for both Model 1 and Model 3.\\

We have chosen the interpretation with the smaller value of the time lags between $|\delta kT_{in}|$, $|\delta \dot H|$ and $|\delta \tau|$, $|\delta \dot H|$ such that the heating rate acts as the driver of the phenomenon. However an alternative picture is also possible where the inner disk temperature will vary earlier than the other variations but then the absolute time lags will be larger than what is reported in Table \ref{tab:model}.\\

While in principle, we could consider more complex models involving changes in the fraction of photons entering the corona $f$ but we defer such detailed analysis for a future work where we will consider a larger number of observations and include the effects of reflection.

\section{DISCUSSION AND CONCLUSION}
\label{sec:results}

In this work, we computed the expected energy dependent fractional r.m.s and time-lag for a system whose time-averaged spectrum can be broadly described as emission from a truncated disk and thermal Comptonization. Comparing the results with data provides information on the physical quantities that are varying to produce the timing phenomenon. We applied the technique to a QPO observed by AstroSat/LAXPC for GRS 1915+105 and found that the observations seem to require that there is variation in the accretion rate and inner radius of the disk along with the heating rate and optical depth of the corona, with time delays between them.\\

While the model used here describes the salient parts of the observed spectrum, it should be emphasised that the technique is general enough to incorporate other important features such as absorption and reflection. Indeed, any spectral decomposition can be used to predict the timing properties of the system, provided the set of spectral parameters used, correspond to physically relevant ones. The only issue is that the parameter space becomes large for complicated models, since for each spectral parameter, its variability amplitude and time lag becomes  parameters for explaining the timing phenomenon.\\

However, there are two important caveats. First the technique does not (and cannot in a straight forward manner) incorporate
light travel time effects and hence it should be applied only to situations where the inferred time-lag  between spectral parameters
is significantly larger than any characteristic light travel time-scale of the system. For the analysis done here the time lag between heating rate of the corona and the inner disk temperature is of the order of 20 milliseconds. The corresponding light travel distance is $6 \times 10^8$ cms which for a black hole mass of $\sim 10 M_\odot$,  would be about $\sim 400$ Gravitational radii. Thus as long as the scale of the system (which may be considered as the inner disk radius) is significantly less than this value, the analysis should be consistent. However, for shorter inferred time scales this may not hold true. Secondly the analysis is a linear one and non-linear effects which would be relevant for large amplitude oscillations are not taken into account. In this work, the QPO and its harmonic are considered separately i.e. the physical parameters are considered to also have variations in both frequencies. However as shown by \citet{Mir et al.{2016}}, it may be possible that the observed harmonic is the non-linear manifestation of variation of physical parameters varying only at the fundamental frequency. To verify this, one would need to  numerically compute the second differentiation of the spectra with respect to a spectral parameter and consider all second order terms, which may arise as combinations of contributions from different spectral parameter variations. Such computations have to be done to ascertain at which
amplitude level these second order effects become important. 

While in this work we show that to some approximation, variations in the inner disk temperature, radius, coronal heating rate and optical depth can explain the energy dependent timing features of the QPO observed in GRS 1915+105, we refrain from making a detailed discussion regarding the parameter values obtained and their physical interpretation. It would be prudent to undertake a more detailed study of a larger number of
observations (perhaps consisting of different types of QPOs) before making inferences. Such a study should also model the spectra from the
simultaneous Soft X-ray telescope on-board  {\it AstroSat} and consider the effect of reflection components such as the Iron line which is detected in the source. Moreover, it will be also interesting to check whether non-linear effects can explain the harmonics observed. Such analysis can also be undertaken for {\it NICER} data which will provide both timing and spectral information in the soft energy band. Simultaneous {\it NICER} and {\it AstroSat} data would provide rich and well constrained information about the nature of the variability of these sources.

\section*{Acknowledgements}

This research work is utilizing ASTROSAT/LAXPC data, available at Indian Space Science Data Centre (ISSDC). The work has made use of softwares provided by High Energy Astrophysics Science Archive Research Center(HEASARC). AG acknowledges the periodic visits to Inter-University Centre for Astronomy and Astrophysics(IUCAA), Pune to carry out the major part of the work. AG thanks UGC, Govt. of India for providing fellowship under the UGC-JRF scheme(Ref. No.: 1449/CSIR NET JUNE 2019).  

\section*{DATA AVAILABILITY}

The data utilized in this article are available at \textit{Astrosat}-ISSDC website (\url{http://astrobrowse.issdc.gov.in/astro\_archive/archive}). The Fortran code underlying the article can be accessed by a reasonable request to the corresponding author. 




\begin{thebibliography}{99}

\bibitem[\protect\citeauthoryear{Agrawal et al.}{2017}]{Agrawal et al.(2017)}
Agrawal P. C., et al., 2017, JApA, \textbf{38}, 30
\bibitem[\protect\citeauthoryear{Axelsson et al.}{2014}]{Axelsson et al.(2014)}
Axelsson M., Done C., Hjalmarsdotter L., 2014, MNRAS, \textbf{438}, 657
\bibitem[\protect\citeauthoryear{Belloni et al.}{2000}]{Belloni et al.(2000)}
Belloni, T., Klein-Wolt, M., Meńdez, M., van der Klis, M., van Paradijs, J., 2000, Astronomy and Astrophysics, \textbf{355}, 271
\bibitem[\protect\citeauthoryear{Blum et al.}{2009}]{Blum et al.(2009)}
Blum, J. L., Miller, J. M., Fabian, A. C., et al. 2009, ApJ, \textbf{706}, 60
\bibitem[\protect\citeauthoryear{Bhargava et al.}{2019}]{Bhargava et al.(2019)}
Bhargava, Yash; Belloni, Tomaso; Bhattacharya, Dipankar; Misra, Ranjeev, 2019, MNRAS, \textbf{488}, 720
\bibitem[\protect\citeauthoryear{Belloni et al.}{2012}]{Belloni et al.(2012)}
Belloni, T. M., Sanna, A., and Méndez, M., 2012, MNRAS, \textbf{426}, 1701
\bibitem[\protect\citeauthoryear{Chakrabarti \& Manickam}{2000}]{Chakrabarti & Manickam(2000)}
Chakrabarti SK, Manickam SG., 2000, Ap. J. \textbf{531},L41
\bibitem[\protect\citeauthoryear{Fragile et al.}{2007}]{Fragile et al.(2007)}
Fragile C. P., Blaes O. M., Anninos P., Salomonson J. D., 2007, ApJ, \textbf{668}, 417
\bibitem[\protect\citeauthoryear{Homan et al.}{2001}]{Homan et al.(2001)}
Homan, J., Wijnands, R., van der Klis, M., Belloni, T., van Paradijs, J., Klein-Wolt, M., Fender, R., and Méndez, M., 2001, ApJ, \textbf{132}, 377
\bibitem[\protect\citeauthoryear{Ingram et al.}{2009}]{Ingram et al.(2009)}
Ingram A., Done C., Fragile P. C., 2009, MNRAS, \textbf{397}, L101
\bibitem[\protect\citeauthoryear{Misra et al.}{2013}]{Misra et al.(2013)}
Misra, R., Mondal, S., ApJ, 2013 \textbf{779}, 71
\bibitem[\protect\citeauthoryear{Mir et al.}{2016}]{Mir et al.{2016}}
Mir, Mubashir Hamid, Misra, Ranjeev, Pahari, Mayukh, Iqbal, Naseer, Ahmad, Naveel, 2016, MNRAS, \textbf{457}, 2999
\bibitem[\protect\citeauthoryear{Motta}{2016}]{Motta S.E.(2016)}
Motta S. E., Astronomische Nachrichten, 2016, \textbf{337}, 398
\bibitem[\protect\citeauthoryear{Bari et al.}{2019}]{Bari et al.(2019)}
Maqbool, Bari, et al., 2019, MNRAS, \textbf{486}, 2964
\bibitem[\protect\citeauthoryear{Nowak et al.}{1999}]{Nowak et al.(1999)}
Nowak M. A., Vaughan B. A., Wilms J., Dove J. B., Begelman M. C., 1999, ApJ, \textbf{510}, 874
\bibitem[\protect\citeauthoryear{Pahari et al.}{2013}]{Pahari et al.(2013)}
Pahari, Mayukh, Neilsen, Joseph, Yadav, J. S., Misra, Ranjeev, Uttley Phil, 2013, ApJ, \textbf{778}, 136
\bibitem[\protect\citeauthoryear{Reig et al.}{2000}]{Reig et al.(2000)}
Reig. P., et al., 2000, ApJ, \textbf{541}, 883
\bibitem[\protect\citeauthoryear{Remillard et al.}{2002}]{Remillard et al.(2002)}
Remillard, R. A., Muno, M. P., McClintock, J. E., and Orosz, J. A., 2002, ApJ, \textbf{580}, 1030
\bibitem[\protect\citeauthoryear{Rodriguez et al.}{2002a}]{Rodriguez et al.(2002a)}
Rodriguez, J., Varnière, P., Tagger, M., \& Durouchoux, P., 2002, Astronomy and Astrophysics, \textbf{387}, 487
\bibitem[\protect\citeauthoryear{Remillard McClintock}{2006}]{Remillard & McClintock(2006)}
Remillard, R. A., \& McClintock, J. E., 2006, ARA\&A, \textbf{44}, 49
\bibitem[\protect\citeauthoryear{Divya et al.}{2019}]{Divya et al.(2019)}
Rawat, Divya, et al., 2019, The Astrophysical Journal, \textbf{870}, 4
\bibitem[\protect\citeauthoryear{Singh et al.}{2016,2017}]{Singh et al.(2016)}
Singh K. P., et al., 2016, SPIE, \textbf{9905}, 99051E
\bibitem[\protect\citeauthoryear{Singh et al.}{2016,2017}]{Singh et al.(2017)}
Singh K. P., et al., 2017, JApA, \textbf{38}, 29
\bibitem[\protect\citeauthoryear{Tagger et al.}{1999}]{Tagger et al.(1999)}
Tagger M, Pellat R, 1999, Astron. Astrophys., \textbf{349}, 1003
\bibitem[\protect\citeauthoryear{Titarchuk et al.}{2000}]{Titarchuk et al.(2000)}
Titarchuk L, Osherovich V , 2000, Ap. J. \textbf{542}, L111
\bibitem[\protect\citeauthoryear{Uttley et al.}{2011}]{Uttley et al.(2011)}
Uttley, P., Wilkinson, T., Cassatella, P., Wilms, J., Pottschmidt, K., Hanke, M., B¨ock, M., 2011, MNRAS, \textbf{414}, L60
\bibitem[\protect\citeauthoryear{Wijnands et al.}{1999}]{Wijnands et al.(1999)}
Wijnands, R., Homan, J., and van der Klis, M., 1999, ApJ, \textbf{526}, L33
\bibitem[\protect\citeauthoryear{Wilkins et al.}{2015}]{Wilkins et al.(2015)}
Wilkins D. R., Gallo L. C., 2015, MNRAS, \textbf{448}, 703
\bibitem[\protect\citeauthoryear{Yadav et al.}{2016}]{Yadav et al.(2016a)}
Yadav J. S., et al., 2016, SPIE, \textbf{9905}, 99051D
\bibitem[\protect\citeauthoryear{Yadav et al.}{2016}]{Yadav et al.(2016b)}
Yadav, J. S., et al., 2016, ApJ, \textbf{833}, 27 
\bibitem[\protect\citeauthoryear{Zdziarski et al.}{1996}]{Zdziarski et al.(1996)}
Zdziarski A. A., Johnson W. N., Magdziarz P., 1996, MNRAS, \textbf{283}, 193
\bibitem[\protect\citeauthoryear{Zycki et al.}{1999}]{Zycki et al.(1999)}
Zycki P. T., Done C., Smith D. A., 1999, MNRAS, \textbf{309}, 561
\end{thebibliography}





\bsp	
\label{lastpage}
\end{document}